\documentclass[aps,prb,twocolumn, 10pt, superscriptaddress]{revtex4-2}
\usepackage{graphicx}
\usepackage{amsmath}
\usepackage{amssymb}
\usepackage{mathtools}
\usepackage{xcolor}
\usepackage{physics}
\usepackage[colorlinks=true, citecolor=blue, urlcolor=blue, linkcolor=blue,bookmarks=false,hypertexnames=true]{hyperref} 
\usepackage[T1]{fontenc} 

\usepackage{soul}

\begin{document}
\graphicspath{}

\title{Theory of excitonic complexes in gated WSe$_2$ quantum dots}

\author{Daniel Miravet}
\thanks{dmiravet@uottawa.ca}
\affiliation{Department of Physics, University of Ottawa, Ottawa, Ontario, K1N 6N5, Canada}

\author{Ludmi\l a Szulakowska}
\affiliation{Department of Physics, University of Ottawa, Ottawa, Ontario, K1N 6N5, Canada}

\author{Maciej Bieniek} 
\affiliation{Institute of Theoretical Physics, Wroc\l aw University of Science and Technology, Wybrze\.ze Wyspia\'nskiego 27, 50-370 Wroc\l aw, Poland}

\author{Marek Korkusi\'nski}
\affiliation{Department of Physics, University of Ottawa, Ottawa, Ontario, K1N 6N5, Canada}
\affiliation{Quantum and Nanotechnologies Research Centre,
National Research Council of Canada, Ottawa, ON, K1A 0R6, Canada}

\author{Pawe\l\ Hawrylak}
\affiliation{Department of Physics, University of Ottawa, Ottawa, Ontario, K1N 6N5, Canada}

\date{\today}

\begin{abstract}
Single-layer quantum dot gate potential causes type-II band alignment, i.e. electrostatically confines holes and repels electrons, or vice versa. Hence, the confinement of excitons in gated type II quantum dots involves a delicate balance of the repulsion of electrons due to the gate potential with the attraction caused by the Coulomb interaction with a hole localized in the quantum dot. This work presents a theory for neutral excitonic complexes within gated $\text{WSe}_\text{2}$ quantum dots, considering spin, valley, electronic orbitals, and many-body interactions.  We analyze how the electron-hole attraction depends on a range of system parameters, such as screened  Coulomb interaction, strength of confinement of holes, and repulsion of electrons. Using an atomistic tight binding model we compute valence and conduction band states within a computational box comprising over one million atoms with applied gate potential. The atomistic wavefunctions are then used to calculate direct and exchange Coulomb matrix elements for a fictitious type I quantum dot, and to obtain a spectrum of interacting electron-hole pairs. Next, we study the effect of repulsive potential, pulling away electrons from the valence hole. We determine whether electron-hole pairs are sufficiently attracted to overcome electron repulsion by the confinement potential. Finally, we compute the dipole transition between hole and electron states to obtain the absorption spectrum.

\end{abstract}

%\keywords{Suggested keywords}%Use showkeys class option if keyword
                              %display desired
\maketitle

%Introduction 
%%%%%%%%%%%%%%%%%%%%%%%%%%%%%%%%%%%%%%%%%%%%%%%%%%%%%%%%%%%%%%%%%%%%%%%%%%%%%%%%%%%%%%%%%%%%%%%
\section{Introduction}
Absorption of a single photon generates a single exciton, comprising an electron-hole pair. Gated quantum dots enable precise manipulation of a controlled number of either electrons or holes within a single layer of 2D materials, such as $\text{WSe}_\text{2}$. However, confining excitons in gated type II quantum dots presents a nontrivial challenge, as the electron is repelled by the gate potential when the hole is confined in the dot. Should a single exciton be confined in a gated quantum dot, its presence will be detected in transport.

Transition metal dichalcogenides (TMDCs) such as $\text{WSe}_\text{2}$ have attracted interest due to their intriguing electronic, optical, and magnetic properties. These arise from strong electron-electron interactions and dimensionality lowering in a single atomic layer.
Recent advancements in TMDCs, fabricated at the few-atomic-layer scale, have expanded our understanding of low-dimensional physics \cite{kadantsev2012electronic,Goh2020,Sierra2021,splendiani2010emerging,roch2019spin,scrace2015magnetoluminescence,van2019virtual,komsa2012two,boddison2021gate,Boddison20231DChannel,mueller2018exciton,schneider2018two,conley2013bandgap,Manzeli_Kis_2017}. Unlike monolayer graphene, TMDCs possess a band gap, enabling optoelectronic devices \cite{CastroNeto_Geim_2009,Manzeli_Kis_2017,Sierra2021,Mak_Heinz_2010}.

TMDCs consist of transition metal atoms bonded to chalcogenide dimers. Their electronic and optical behavior predominantly stems from the $d$ orbitals of metal atoms \cite{altintacs2021spin,bieniek2020effect,bieniek2018band,MiravetPRB2023holesQD}. The strong localization of these orbitals, combined with reduced screening in two dimensions, gives rise to pronounced correlation effects, ultimately leading to the formation of excitons with large binding energies \cite{OtsukaCriticalityMott2016,BieniekExcitons2022,BorgesexcitonQ2023,WuExcitonMoS2_2015,Goryca2019,Thureja2022}.

The spin-orbit coupling in TMDCs gives rise to distinct spin-dependent phenomena. Notably, $\text{WSe}_\text{2}$ stands out due to its substantial spin-orbit splitting (approximately $500$ meV) in the valence band, distinguishing it from other TMDC materials \cite{le2015spin,alidoust2014observation}. Additionally, it is worth noting that while the bottom of the conduction band in valleys K and -K is predominantly composed of $m_d = 0$ orbitals, the top of the valence band in valleys K and -K primarily comprises $m_d = -2$ and $m_d = 2$ orbitals, respectively, indicating orbital asymmetry between the valleys \cite{bieniek2020effect}.

Recent progress in graphene and TMDC materials, combined with lateral gating techniques \cite{Ciorga2000}, has enabled the fabrication of lateral gated quantum dots (QDs) \cite{Guclu_Hawrylak_2014,Guttinger_Ensslin_2010,McGuire_2016,Wang_Ruffieux_2017,Wang_Kim_2018,Pisoni_Ensslin_2018,Volk_Stampfer_2011,Allen_Yacoby_2012,Eich_Ensslin_2018,Kurzmann_Ihn_2019,freitag2016electrostatically,YasserGrapheneQD2023,Davari2020}. By applying lateral metallic gates, electrons or holes can be laterally electrostatically confined, resulting in the creation of atomic-like states \cite{altintacs2021spin,bieniek2020effect, Eich_Ensslin_2018,freitag2016electrostatically,Kurzmann_Ihn_2019,BrotonsGisbert_Gerardot_2019,Lu_Srivastava_2019,Chakraborty_Vamivakas_2018,Zhang_Guo_2017,bhandari2018imaging,chen2018magnetic,LudkaQD2020,boddison2021gate,JarekPRA2021} and strongly interacting many-particle complexes \cite{Marek_nanoletters_2023,MiravetPRB2023holesQD,pawłowski2024interacting}.

 Here we present a theory of excitons \cite{HePRL2014,WangPRL2015}  in gated $\text{WSe}_\text{2}$ quantum dots \cite{altintacs2021spin,MiravetPRB2023holesQD}, and predict a set of parameters involving spin, valley, topology, and many-body interactions for which exciton can be created and confined in a type II gated quantum dot \cite{boddison2021gate,Pisoni_Ensslin_2018,Davari2020,BrotonsGisbert_Gerardot_2019,Lu_Srivastava_2019,Zhang_Guo_2017,song2015gate}.
 We describe our quantum dot within a computational box containing over a million atoms, employing a tight-binding microscopic Hamiltonian \cite{BieniekExcitons2022}. The gated quantum dot is defined by applying lateral gates that result in a  confining potential, which is attractive for holes and repulsive for electrons. By expanding the quantum dot wavefunction in bulk band states, we obtain quantum dot energy levels and wavefunctions, capturing the effects of confinement potential, size, valley, and spin \cite{MiravetPRB2023holesQD,Boddison20231DChannel}. We then construct the basis of electron-hole pairs and expand the exciton wavefunction in terms of these pairs. Using atomistic wavefunctions, we calculate Coulomb matrix elements important for determining whether electron-hole pairs are sufficiently attracted to overcome electron repulsion by the confining potential. We then obtain exciton levels in an auxiliary quantum dot, which confines electrons and holes. Expanding the potential repelling an electron in a quantum dot in the basis of exciton states in the auxiliary quantum dot allows for the electron's attraction to the hole localized in the dot, overcoming the repulsive potential of the gate. Finally, we determine the confining potential and strength/screening of electron-electron interactions for which exciton is confined in a quantum dot and can be detected in transport experiments.

The paper is organized as follows. In Section \ref{sect:model}, we describe the single-particle states in the gated $\text{WSe}_\text{2}$ quantum dot, along with an overview of the formulation of the many-body problem and the computation of the dipole matrix elements. The phase diagram describing stable exciton as a function of confining potential strength and Coulomb interactions and resulting excitonic absorption spectrum are presented in  Section \ref{sect:discussion}.
The paper concludes with a summary and key findings in Section \ref{sect:conclusions}.
%%%%%%%%%%%%%%%%%%%%%%%%%%%%%%%%%%%%%%%%%%%%%%%%%%%%%%%%%%%%%%%%%%%%%%%%%%%%%%%%%%%%%%%%%%%%%%%%

%%%%%%%%%%%%%%%%%%%%%%%%%%%%%%%%%%%%%%%%%%%%%%%%%%%%%%%%%%%%%%%%%%%%%%%%%%%%%%%%%%%%%%%%%%%%%%%%
% Model
\section{MODEL}
\label{sect:model}

%%%%%%%%%%%%%%%%%%%%%%%%%%%%%%%%%%%%%%%%%%%%%%%%%%%%%%%%%%%%%%%%%%%%%
%%%%%%%%%%%%%%%%%%%%%%%%%%%%%%%%%%%%%%%%%%%%%%%%%%%%%%%%%%%%%%%%%%%%%%%%%%%%%%%%%%%%%%%%%%%%%%%%
\subsection{Tight-Binding model}
\label{sect:TB model}
We construct our computational box consisting of W and Se atoms arranged in a rhombus. By applying periodic boundary conditions, we obtain a set of allowed wavevectors $\vec{k}$.  The wavefunctions of the wavefunctions of computational rhombus at each wavevector $\vec{k}$ are constructed as a linear combination of Bloch functions on the tungsten and selenium atoms sublattices. In particular, we consider six such Bloch functions indexed by $l$ ($l=1,..,6$):
%----------------------------------------- Bulk wavefunction -----------------------------------------
\begin{equation}
\ket{\phi^{p}_{\vec{k}\sigma}} = \sum_{l=1}^{6} A^{p}_{\vec{k}\sigma, l} \ket{\phi^{\textrm{even}}_{\vec {k},l}}\otimes\ket{\chi_{\sigma}},
\label{eq:4}
\end{equation}
%-----------------------------------------
where $\ket{\chi_{\sigma}}$ represents the spinor part of the wavefunction and
%-----------------------------------------
\begin{equation}
\ket{\phi^{\textrm{even}}_{\vec{k},l}}=\frac{1}{\sqrt{N_{\textrm{UC}}}} \sum_{\vec{R_{l}}=1}^{N_{\textrm{UC}}} e^{i\vec{k}\cdot\vec{R_{l}}} \varphi^{\textrm{even}}_{l}\left(\vec{r}-\vec{R_{l}}\right),
\end{equation}
%-----------------------------------------
are  Bloch functions constructed with orbitals $\varphi^{\textrm{even}}_{l}$, which are even with respect to the metal plane. Here, $N_{\textrm{UC}}$ denotes the number of unit cells, and $\vec{R}_l$ defines the position of orbitals in the computational box. By diagonalizing the $6\times 6$ bulk Hamiltonian \cite{BieniekExcitons2022}, we obtain the even bulk energy bands $E_{k\sigma}^{\textrm{p}}$ and wavefunctions $\ket{\phi^{\textrm{p}}_{\vec{k}\sigma}}$. 

%-----------------------  fig1 ------------------------
\begin{figure}[ht]
\centering          
\includegraphics[width=9cm, height=7cm]{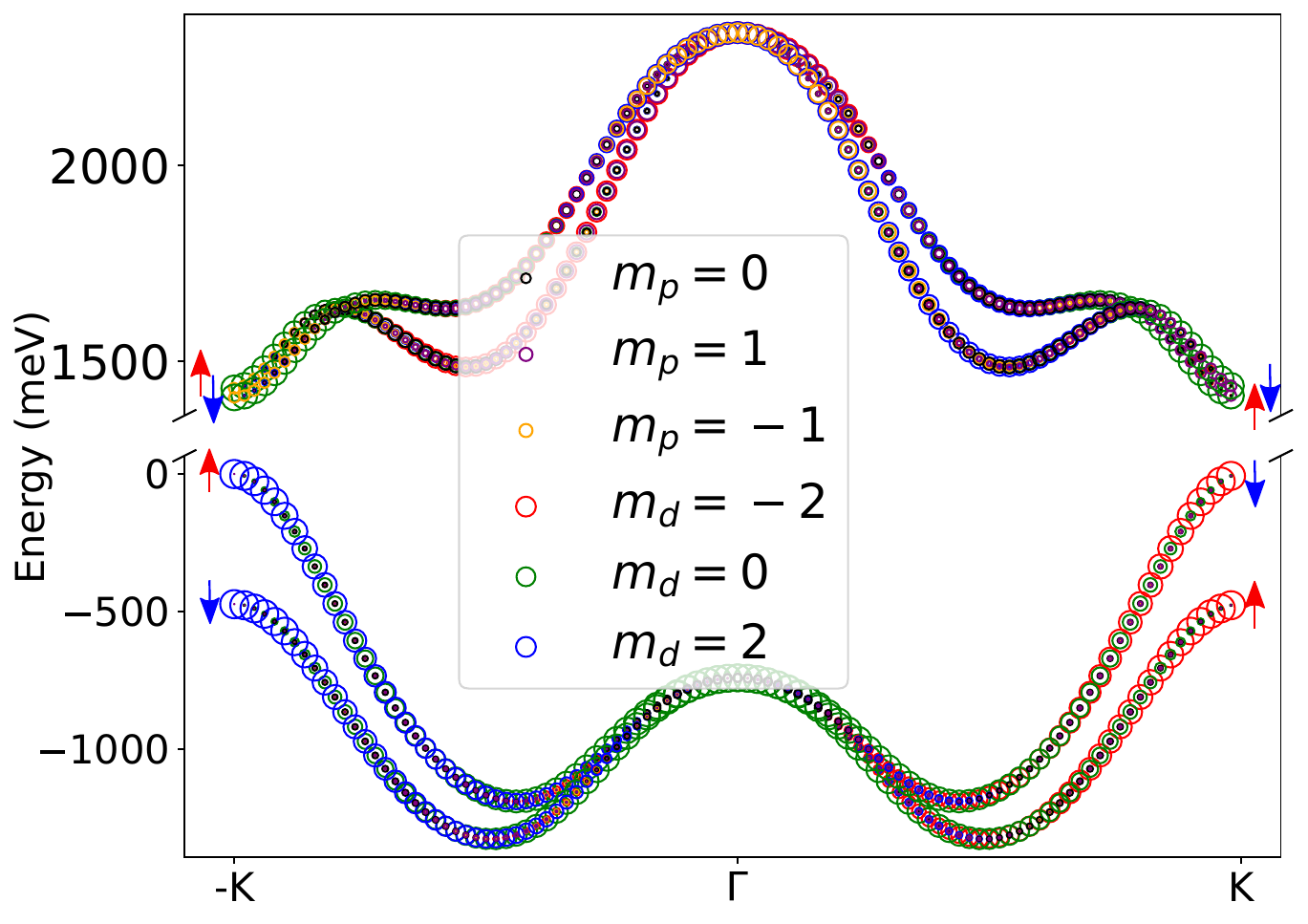}
\caption{
(Color online)   Highest energy valence bands and lowest energy conduction bands as a function of wavevector $k$ over the path $-K \rightarrow \Gamma \rightarrow +K $. For the valence band, the states in the $+K$ valley are composed mainly of $m_{d} = -2$ orbitals, while states in the $-K$ valley are composed primarily of $m_{d} = +2$ orbitals. On the other hand, the states in the conduction band $+K$ valley are composed primarily of $m_{d} = 0$ and $m_{p} = +1$ orbitals while states in the $-K$ valley are composed primarily of $m_{d} = 0$ and $m_{p} = -1$ orbitals. The $+K$ and $-K$ points represent the global valley maxima (minima) of the valence (conduction) band. } 
\label{fig1}
\end{figure}
%-----------------------------------------
Figure \ref{fig1} shows the energy $E_{k\sigma}^{p}$ of the highest even valence band (VB) and lowest even conduction band (CB)  along the path $-K\rightarrow \Gamma \rightarrow K $. It is important to note that the spin splitting $\Delta_{SOC}$ is opposite in these valleys, with a magnitude of approximately $470$ meV \cite{le2015spin} for the valence band and $50$ meV for the conduction band. While the bottom of the conduction band in valleys K and -K is mainly comprised of $m_d = 0$ orbitals, the top of the valence band in valleys K and -K consists primarily of $m_d = -2$ and $m_d = 2$ orbitals, respectively. 

%%%%%%%%%%%%%%%%%%%%%%%%%%%%%%%%%%%%%%%%%%%%%%%%%%%%%%%%%%%%%%%%%%%%%%%%%%%%%%%%%%%%%%%%%%%%%%%%
\subsection{Gate-Defined Quantum Dot}
\label{sect:GateQD}
In Ref. \cite{MiravetPRB2023holesQD}, it was shown that by applying a confining Gaussian potential due to lateral gates $V_{QD}(\vec{r})=V_{0}\exp\left(-r^{2}/R_{QD}^2 \right)$, where $V_0>0$, to a single layer of $\text{WSe}_\text{2}$, a quantum dot with confined hole states can be created, with $R_{QD}$ denoting the QD radius. The single-particle Hamiltonian describing a particle in a $\text{WSe}_\text{2}$ QD can be expressed as the sum of the bulk computational box Hamiltonian $H_{\textrm{TB}}$ and the confining potential $V_{QD}$ \cite{altintacs2021spin}. The wavefunction $|\Phi^{s}\rangle$, associated with the state $s$, can be determined by solving the Schr\"odinger  equation:
    %----------------------------------------- Hamiltonian basic form -----------------------------------------
\begin{equation}
 (H_{\textrm{TB}}+V_{QD}(\vec{r})) \ket{\Phi^{s}}= E^s \ket{\Phi^{s}}.
\label{main:hamqdeq1}
\end{equation}
    %-----------------------------------------

We  expand QD wavefunction $\ket{\Phi^{s}}$  in terms of bulk states of the computational box, given by Eq. \ref{eq:4}:
    %----------------------------------------- QD wavefunction -----------------------------------------
\begin{equation}
\ket{\Phi^{s}}= \sum_{\vec{k}} \sum_p \sum_{\sigma} B^{\textrm{s,p}}_{\vec{k}\sigma} \ket{\phi^p_{\vec{k}\sigma}},
\label{eq:KetQD}
\end{equation}
    %-----------------------------------------
where  $p$ runs over the highest energy even valence band and the lowest energy even conduction band. Since we focus on the highest valence band and lowest conduction band states, we apply a cutoff to the $\vec{k}$ points near $K$ and $-K$ valleys by selecting vectors $\vec{k}$ that satisfy the condition $|\vec{k}\pm \vec{K}| < \eta |\vec{K}|$. This approach helps us focus on the reciprocal space's relevant region. In this study, we have maintained a fixed value of $\eta = 0.1$ throughout our calculations \cite{MiravetPRB2023holesQD}.

Then, expanding the QD Hamiltonian in terms of the band states, we have

    %----------------------------------------- Splited Hamiltonian -----------------------------------------
\begin{align}
\hat{H} =& \hat{H}_{\textrm{TB}} + \sum_{\vec{k},\vec{q},\sigma} c^{\textrm{VB}^\dagger}_{\vec{k},\sigma} c^{\textrm{VB}}_{\vec{q},\sigma}\bra{\phi^{VB}_{\vec{k}\sigma}}\hat{V}_{QD}\ket{\phi^{VB}_{\vec{q}\sigma}} \nonumber \\
 &+ \sum_{\vec{k},\vec{q},\sigma} c^{\textrm{CB}^\dagger}_{\vec{k},\sigma}c^{\textrm{CB}}_{\vec{q},\sigma}\bra{\phi^{\textrm{CB}}_{\vec{k}\sigma}}\hat{V}_{QD}\ket{\phi^{\textrm{CB}}_{\vec{q}\sigma}}\nonumber \\
  &+ \sum_{\vec{k},\vec{q},\sigma} c^{\textrm{VB}^\dagger}_{\vec{k},\sigma}c^{\textrm{CB}}_{\vec{q},\sigma}\bra{\phi^{\textrm{VB}}_{\vec{k}\sigma}}\hat{V}_{QD}\ket{\phi^{\textrm{CB}}_{\vec{q}\sigma}} + h.c  .
\end{align}
    %-----------------------------------------

Here, the first term, diagonal, corresponds to the CB and VB energies of the tight-binding computational box Hamiltonian. The second and third terms denote the interaction between the quantum dot confining potential and the valence and conduction band states, respectively. The final term represents the coupling between conduction and valence band states induced by the confining potential.

\begin{figure}[ht]
\centering          
\includegraphics[width=9cm, height=4.15cm]{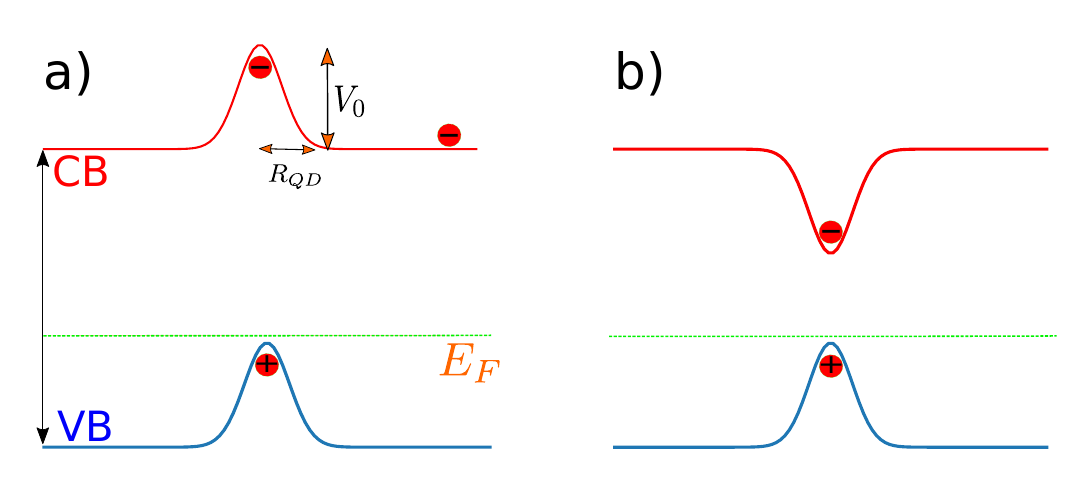}
\caption{
(Color online)  (a) Schematic representation of the confinement in the QD created by a gate potential. Confined states in the valence band are pushed into the gap while states in the conduction band are pushed away from the gap. b) Schematic representation of the auxiliary QD. The new potential confines both hole and electron states. } 
\label{fig2}
\end{figure}

Figure \ref{fig2} schematically illustrates the effect of the confining potential. We can deplete some confined valence band states by appropriately tuning the chemical potential or equivalently creating confined hole QD states. While the confined hole states are pushed into the band gap, the low energy conduction band states are pushed away from the center of the quantum dot and delocalized. As a consequence, the type-II-like electron-hole pairs are the most energetically favorable - with the hole confined within the QD and the electron outside of it. However, when Coulomb interaction is included, this picture changes due to the strong electron-hole attraction, which binds the electron to the hole in a type I exciton state. This exciton is also strongly coupled to light because both electron and hole states are confined inside the QD. Therefore, we will focus on determining the parameters of the quantum dot, the strength of the potential repelling an electron vs the strength of attraction by the hole in the formation of the type I exciton bound in a quantum dot.

To artificially select  CB states localized in the quantum dot  (see Figure \ref{fig2}b), we subtract and add the term $ \sum_{\vec{k},\vec{q},\sigma} c^{\textrm{CB}^\dagger}_{\vec{k},\sigma}c^{\textrm{CB}}_{\vec{q},\sigma}\bra{\phi^{\textrm{CB}}_{\vec{k}\sigma}}\hat{V}_{QD}\ket{\phi^{\textrm{CB}}_{\vec{q}\sigma}}$ in the Hamiltonian (Eq 5)
confining and repelling electrons from the quantum dot:

     %----------------------------------------- Splited Hamiltonian 2-----------------------------------------
\begin{align}
\hat{H} =& \hat{H}_{\textrm{TB}} + \sum_{\vec{k},\vec{q},\sigma} c^{\textrm{VB}^\dagger}_{\vec{k},\sigma}c^{\textrm{VB}}_{\vec{q},\sigma}\bra{\phi^{VB}_{\vec{k}\sigma}}\hat{V}_{QD}\ket{\phi^{VB}_{\vec{q}\sigma}} \nonumber \\
 &- \sum_{\vec{k},\vec{q},\sigma} c^{\textrm{CB}^\dagger}_{\vec{k},\sigma}c^{\textrm{CB}}_{\vec{q},\sigma}\bra{\phi^{\textrm{CB}}_{\vec{k}\sigma}}\hat{V}_{QD}\ket{\phi^{\textrm{CB}}_{\vec{q}\sigma}}\nonumber \\
   &+ \sum_{\vec{k},\vec{q},\sigma} c^{\textrm{VB}^\dagger}_{\vec{k},\sigma}c^{\textrm{CB}}_{\vec{q},\sigma}\bra{\phi^{\textrm{VB}}_{\vec{k}\sigma}}\hat{V}_{QD}\ket{\phi^{\textrm{CB}}_{\vec{q}\sigma}} + h.c   \nonumber \\
   &+ 2\sum_{\vec{k},\vec{q},\sigma} c^{\textrm{CB}^\dagger}_{\vec{k},\sigma}c^{\textrm{CB}}_{\vec{q},\sigma}\bra{\phi^{\textrm{CB}}_{\vec{k}\sigma}}\hat{V}_{QD}\ket{\phi^{\textrm{CB}}_{\vec{q}\sigma}}.
  \end{align}
    %-----------------------------------------
The first three lines of this Hamiltonian describe an auxiliary QD that confines both electrons and holes. while the last line is the correction (perturbation), which pulls only the electrons in the CB  away from the quantum dot.  We first obtain the single-particle states for auxiliary QD confining both holes and electrons,  and then compute confined exciton states by diagonalizing the many-body Hamiltonian in Sec. \ref{sect:QD excitonic spectrum}. This allows us to determine whether the repulsive electron potential can pull the electron from the bound exciton state:
  %----------------------------------------- Splited Hamiltonian 3-----------------------------------------
\begin{align}
\hat{H} =& \hat{H}_{\textrm{aux}}  + \hat{V}_\textrm{corr}.
\label{eq:QDHamilSplited}
\end{align}
Here
%----------------------------------------- Splited Hamiltonian 3-----------------------------------------
\begin{align}
\hat{H}_{\textrm{aux}} =&\hat{H}_{\textrm{TB}} + \sum_{\vec{k},\vec{q},\sigma} c^{\textrm{VB}^\dagger}_{\vec{k},\sigma}c^{\textrm{VB}}_{\vec{q},\sigma}\bra{\phi^{VB}_{\vec{k}\sigma}}\hat{V}_{QD}\ket{\phi^{VB}_{\vec{q}\sigma}} \nonumber \\
 &- \sum_{\vec{k},\vec{q},\sigma} c^{\textrm{CB}^\dagger}_{\vec{k},\sigma}c^{\textrm{CB}}_{\vec{q},\sigma}\bra{\phi^{\textrm{CB}}_{\vec{k}\sigma}}\hat{V}_{QD}\ket{\phi^{\textrm{CB}}_{\vec{q}\sigma}}\nonumber \\
  &+ \sum_{\vec{k},\vec{q},\sigma} c^{\textrm{VB}^\dagger}_{\vec{k},\sigma}c^{\textrm{CB}}_{\vec{q},\sigma}\bra{\phi^{\textrm{VB}}_{\vec{k}\sigma}}\hat{V}_{QD}\ket{\phi^{\textrm{CB}}_{\vec{q}\sigma}} + h.c ,
\label{Eq.AuxQDHam}
\end{align}
is the Hamiltonian for the auxiliary QD which confines electrons and holes, and

 %----------------------------------------- V correction-----------------------------------------
\begin{align}
\hat{V}_\textrm{corr} =&  2\sum_{\vec{k},\vec{q},\sigma} c^{\textrm{CB}^\dagger}_{\vec{k},\sigma}c^{\textrm{CB}}_{\vec{q},\sigma}\bra{\phi^{\textrm{CB}}_{\vec{k}\sigma}}\hat{V}_{QD}\ket{\phi^{\textrm{CB}}_{\vec{q}\sigma}},
\label{eq:Vcorr}
\end{align}
is a correction that pulls the electron in CB away from the hole in a quantum dot. In section \ref{sect:Auxiliary QD} we will determine the spectrum of this auxiliary QD as the first step toward the understanding of the original QD.

In numerical calculations, we employ a computational box containing $N_1 \times N_2 = 633 \times 633$ unit cells, corresponding to $3 \cdot N_1 \times N_2 =$ 1,202,067 atoms. Periodic boundary conditions are enforced on the computational box, resulting in a discrete set of $\vec{k}$ points in reciprocal space.
%%%%%%%%%%%%%%%%%%%%%%%%%%%%%%%%%%%%%%%%%%%%%%%%%%%%%%%%%%%%%%%%%%%%%%%%%%%%%%%%%

%%%%%%%%%%%%%%%%%%%%%%%%%%%%%%%%%%%% QD spectrum %%%%%%%%%%%%%%%%%%%%%%%
%-----------------------  fig2 ------------------------
%-----------------------------------------

%%%%%%%%%%%%%%%%%%%%%%%%%%%%%%%%%%%%%%%%%%%%%%%%%%%%%%%%%%%%%%%%%%%%%%%%%%%%%%%%%%%%%%%%%%%%%%%%

% Interactions
%%%%%%%%%%%%%%%%%%%%%%%%%%%%%%%%%%%%%%%%%%%%%%%%%%%%%%%%%%%%%%%%%%%%%%%%%%%%%%%%%%%%%%%%%%%%%%%%
\subsection{Interactions}
\label{sect:interactions}
The optical properties of quantum dots are governed by excitonic complexes. We start here by computing excitons in the auxiliary quantum dot, which requires the diagonalization of the many-body Hamiltonian in the space of electron-hole excitations:
    %---------------------------------- Many Body Hamiltonian --------------------------------------------
\begin{align}
H=&\sum_{i} E^i c^\dagger_i c_i + \frac{1}{2}\sum_{i,j,k,l} \bra{ij}V\ket{kl} c^\dagger_i c^\dagger_j c_k c_l \nonumber \\
&-\sum_{i,j}\sum_{m\in \textrm{occup}} \bra{im}V\ket{mj}c^\dagger_i c_j.
%+ \sum_{i,j}\tau_{i,j}c^\dagger_i c_j.
\label{eq:manybodyHam}
\end{align}
    %-----------------------------------------------------------------------------------------
Here $E^i$  represents the energies of the auxiliary QD single-particle states, obtained by diagonalization of the Hamiltonian given by Eq. \ref{Eq.AuxQDHam}. The operator $c^\dagger_i $ ($c_i$) creates (annihilates) a particle in the auxiliary  QD state $i$. The Coulomb matrix elements, denoted as $\bra{ij}V\ket{kl}$, are expressed in terms of the auxiliary QD single-particle orbitals as 

    %---------------------------------- Vee -----------------------------------
\begin{align}
    \bra{ij}V\ket{kl} =&\int d\vec{r}_{1}\int d\vec{r}_{2}\Phi^{i}(\vec{r}_{1})^{\ast}\Phi^{j}(\vec{r}_{2})^{\ast} V(\vec{r}_{2}-\vec{r}_{1})\nonumber\\
    &\times \Phi^{k}(\vec{r}_{2})\Phi^{l}(\vec{r}_{1}),
\end{align}
    %-----------------------------------------------------------------------------------------
where $\Phi^{i}(\vec{r})=\bra{\vec{r}}\ket{\Phi^{i}}$, and $V_C$ is the Coulomb potential $V_C=e^2/4\pi\epsilon \epsilon_0 |\vec{r}_1-\vec{r}_2|$, with $e$ being the electron charge, $\epsilon_0$ being the vacuum dielectric permittivity and $ \epsilon $ being the dielectric constant. The last term in Eq. \ref{eq:manybodyHam} represents the electron's interaction with the compensating background.

We approximate the ground state of the auxiliary quantum dot as a single Slater determinant with all valence band states occupied, denoted as $\ket{GS}=\Pi_{p\in \textrm{VB}}c_p^\dagger\ket{0}$. Subsequently, we construct the many-body excitations  by generating all possible configurations of electron-hole pairs excited from the ground state with $S_z=0$
\begin{align}
    \ket{X^\mu_\textrm{aux}} =&\sum_{p\in \textrm{VB},q \in \textrm{CB},\sigma_p=\sigma_q} A_{p,q}^\mu c^\dagger_q c_p\ket{\textrm{GS}}.
    \label{eq:excitonGS}
\end{align}
Here, the index $\mu$ enumerates the excitonic states, and $A_{p,q}^\alpha$ represents the expansion coefficients of the corresponding eigenvector.
%%%%%%%%%%%%%%%%%%%%%%%%%%%%%%%%%%%%%%%%%%%%%%%%%%%%%%%%%%%%%%%%%%%%%%%%%%%%%%%%%%%%%%%%%%%%%%%%

%%% exciton in a qdot
%%%%%%%%%%%%%%%%%%%%%%%%%%%%%%%%%%%%%%%%%%%%%%%%%%%%%%%%%%%%%%%%%%%%%%%%%%%%%%%%%%%%%%%%%%%%%%%%
\subsection{Exciton in a quantum dot}
We now have a basis of excitonic states in auxiliary quantum dot and perturbation $\hat{V}_{\textrm{corr}}$ which pulls the electron in CB away from the hole in a quantum dot. The exciton spectrum $\ket{X^\alpha}$ of a gated quantum dot can now be expanded in the basis of exciton states $\ket{X^\mu_\textrm{aux}}$ of the auxiliary QD
\begin{align}
    \ket{X^\alpha} =&\sum_{\mu} D^{\alpha}_{\mu}\ket{ X^\mu_\textrm{aux}}.
    \label{eq:excitonQD}
\end{align}
Then, the quantum dot exciton states satisfy the equation

\begin{align}
    E_{\mu}^{\textrm{aux}} D^{\alpha}_{\mu} + &\sum_{\mu'} \bra{X^{\mu}_\textrm{aux}}
    \hat{V}_\textrm{corr}\ket{X^{\mu'}_\textrm{aux}} D^{\alpha}_{\mu'} = E_{\alpha} D^{\alpha}_{\mu}.
    \label{eq:excitonQDCorrection}
\end{align}

Here $E_{\alpha}^{\textrm{aux}}$ are auxiliary exciton energies and $\bra{X^{\mu}_\textrm{aux}}
    \hat{V}_\textrm{corr}\ket{X^{\mu'}_\textrm{aux}}$ is the effect of repulsive potential acting on electrons in CB only, pulling electrons away from the exciton levels localized in the quantum dot, with a hole attracting the electron.  

%%%%%%%%%%%%%%%%%%%%%%%%%%%%%%%%%%%%%%%%%%%%%%%%

% Absorption spectrum
%%%%%%%%%%%%%%%%%%%%%%%%%%%%%%%%%%%%%%%%%%%%%%%%%%%%%%%%%%%%%%%%%%%%%%%%%%%%%%%%%%%%%%%%%%%%%%%%
\subsection{Absorption spectrum}
\label{sect:Absorption_spectrum}
To understand which excitonic states are optically active we compute the absorption spectrum. The absorption spectrum $A(\omega)$ of an auxiliary quantum dot as a function of the photon energy is computed using Fermi's golden rule as
    %---------------------------------- A(w) -----------------------------------
\begin{equation}
A(\omega)=\sum_{\mu} \left|\bra{X^\mu}P^\dagger\ket{\textrm{GS}} \right|^2 \delta(\omega- E_\mu + E_{\textrm{GS}}),
\end{equation}
    %-----------------------------------------------------------------------------------------
where $E_{\textrm{GS}}$ is the energy of the initial state and $E_\mu$ is the energy of the final excitonic state. The polarization operator $P^\dagger=\sum_{p,q}\delta_{\sigma_p,\sigma_q}\vec{\varepsilon}_0\cdot\vec{D}_{qp}c^\dagger_qc_p$ creates a single pair excitation, $\vec{\varepsilon}_0$ denotes the polarization vector of the circularly polarized light, and the dipole elements $\vec{D}_{qp}$ are given by 

    %---------------------------------- D_pq -----------------------------------
\begin{align}
\vec{D}_{qp}&=\bra{q}\vec{r}\ket{p}\nonumber\\ 
&=\delta_{\sigma_p,\sigma_q} \sum_{\vec{R}_1,\vec{R}_2}\sum_{\vec{l}_1,\vec{l}_2} C^{q\ast}_{\vec{R}_1,l_1}C^{p}_{\vec{R}_2,l_2}\bra{\varphi^{\textrm{even}}_{\vec{R}_1,l_1}}\vec{r}\ket{ \varphi^{\textrm{even}}_{\vec{R}_2,l_2}},
\end{align}
    %-----------------------------------------------------------------------------------------
 where $C^{s}_{\vec{R},l} = \sum_{\vec{k}}A^{p}_{\vec{k}\sigma, l}  B^{\textrm{s,p}}_{\vec{k}\sigma}e^{i\vec{k}\cdot \vec{R}_l}$, and 
    %---------------------------------- D_ij -----------------------------------
\begin{align}
\bra{\varphi^{\textrm{even}}_{\vec{R}_1,l_1}}\vec{r}\ket{ \varphi^{\textrm{even}}_{\vec{R}_2,l_2}} =& \int d\vec{r}\varphi^{\textrm{even}\ast}_{\vec{R}_1,l_1} (\vec{r}-\vec{R}_1-\vec{\tau}_{l_1})\vec{r}  \nonumber\\
&\times \varphi^{\textrm{even}}_{\vec{R}_2,l_2}(\vec{r}-\vec{R}_2-\vec{\tau}_{l_2}).
\end{align}
    %-----------------------------------------------------------------------------------------
%%%%%%%%%%%%%%%%%%%%%%%%%%%%%%%%%%%%%

% Results
\section{Discussion}
\label{sect:discussion}

% Auxiliary QD
%%%%%%%%%%%%%%%%%%%%%%%%%%%%%%%%%%%%%%%%%%%%%%%%%%%%%%%%%%%%%%%%%%%%%%%%%%%%%%%%%%%%%%%%%%%%%%%%
\subsection{Excitons in auxiliary QD}
\label{sect:Auxiliary QD}
\begin{figure*}[ht]
\centering          
\includegraphics[width=\textwidth]{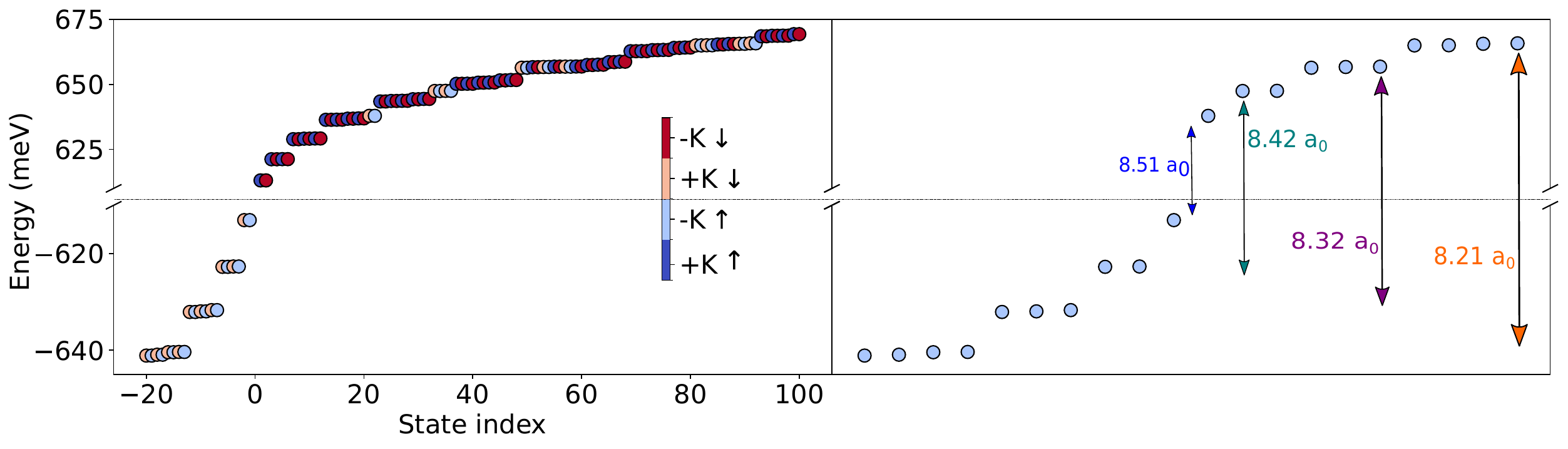}
\caption{
(Color online) 
Auxiliary QD energy levels for $R_{QD}=20$ nm and $V_0=100$ meV.
a) The auxiliary quantum dot energy levels.
b) The highest valence quantum dot states and the conduction QD states with spin-up, localized in the -K valley. The arrows represent the primary allowed transitions between different valence and conduction levels of the QD, the numbers represent the value of the dipole matrix elements of the transitions. The horizontal dashed line represents the Fermi level.} 
\label{fig: AuxQDSP}
\end{figure*}

We obtain the energy levels and wavefunctions of the auxiliary quantum dot by solving Eq. (\ref{Eq.AuxQDHam}). In Figure \ref{fig: AuxQDSP}a, we illustrate the one-particle levels for the QD that confines both electrons and holes. Each color represents different valley and spin combinations. The plot shows that the energy levels exhibit a distinctive grouping into shells, resembling the spectrum of a harmonic oscillator. Due to the significant spin-orbit coupling in the valence band of $\textrm{WSe}_2$, the QD valence band states exhibit spin locking, with spin-up states localized in the -K valley and spin-down states in the +K valley. Conversely, the spin-orbit coupling for the conduction band is relatively small, allowing for all spin and valley combinations within energy windows of $60$ meV.

A consequence of the spin locking is that the lowest transitions in the QD are dark since the closest states in the conduction band are in opposite valleys or with opposite spin. The possible optically active transitions are to spin-up states localized in the -K valley and spin-down states in the +K valley. Figure \ref{fig: AuxQDSP}b shows the highest valence QD states and the conduction QD state sharing the same spin for valley -K. The arrows in Figure \ref{fig: AuxQDSP}a represent the primary allowed transitions between the QD valence and conduction levels. The largest transitions occur between states with the same shell structure, such as s to s or p to p, etc., while dipole transitions between states in different shells are approximately one order of magnitude smaller.

%-----------------------  fig Aux QD SP levels ------------------------
\begin{figure}[ht]
\centering 
\includegraphics[width=9cm, height=7cm]{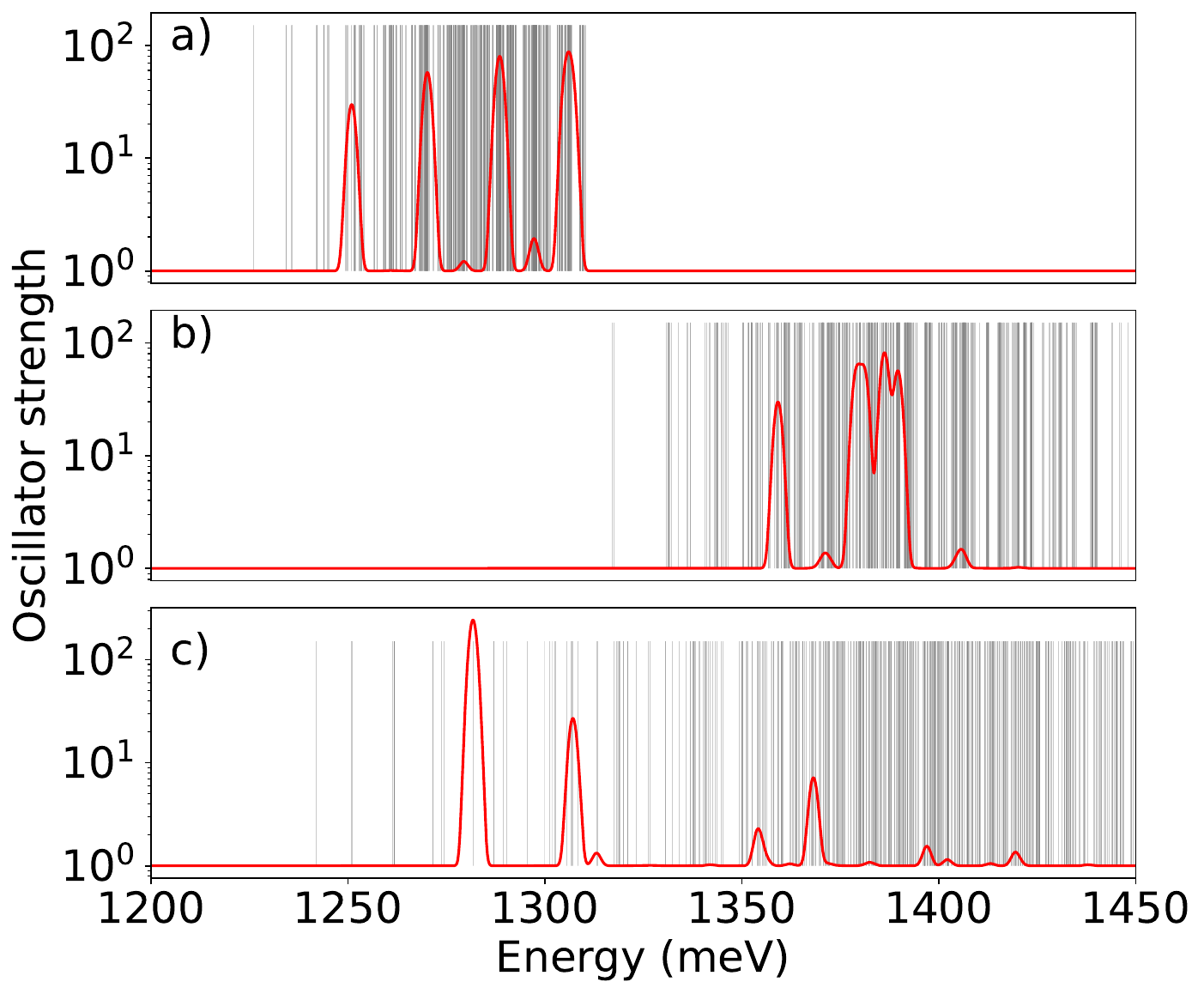}
\caption{
(Color online)  Low energy absorption spectrum for the auxiliary QD with $R_{QD}=20$ nm , $V_0=100$ meV and $\epsilon = 3.9$. Panel a) shows the single-particle picture, b) includes self-energy and vertex corrections on top of it, and c) also allows for the scattering of configurations. The broadening has been added by hand.} 
\label{fig: AuxQD_absorption}
\end{figure}

%-----------------------------------------
In Figure \ref{fig: AuxQD_absorption}, we depict the absorption spectrum as a function of photon energy for the auxiliary quantum dot, that confines both electrons and holes.  The vertical black lines correspond to energies of excitonic states, while the red line represents the absorption strength for the optically active states.

%-----------------------  fig Aux QD absorption ------------------------
%-----------------------------------------

The upper panel illustrates the non-interacting exciton spectrum, which can be understood with the help of Figure \ref{fig: AuxQDSP}. The lowest exciton state is dark, involving transitions between valence and conduction states in the s shells. We observe four peaks with increasing heights, representing transitions from valence states within equivalent shells to conduction states. The peak heights increase for higher energy transitions due to the increased degeneracy of the shells. Smaller peaks represent inter-shell transitions with smaller dipole elements.

Moving to the center panels, we incorporate the electron-hole interaction effect, without considering scattering between different excitonic configurations. The peaks shift blue to higher energies due to the exchange self-energy compensated partially by electron-hole attraction.

The lower panel in Figure \ref{fig: AuxQD_absorption} displays the exciton absorption spectrum, including all terms in the electron-electron interaction matrix elements. Here, we observe a more abrupt distribution of peak heights, given by correlations between different excitonic configurations. Additionally, the brightest peak experiences a redshift to lower energies. The lowest energy peak corresponds to a correlated state, wherein the bright transition between s to s shell contributes approximately $30\%$.

%%%%%%%%%%%%%%%%%%%%%%%%%%%%%%%%%%%%%%%%%%%%%%%%%%%%%%%%%%%%%%%%%%%%%

% Excitons in QD
%%%%%%%%%%%%%%%%%%%%%%%%%%%%%%%%%%%%%%%%%%%%%%%%%%%%%%%%%%%%%%%%%%%%%%%%%%%%%%%%%%%%%%%%%%%%%%%%
\subsection{QD excitonic spectrum}
\label{sect:QD excitonic spectrum}
Now, we will discuss the original QD with a confining potential that confines holes but repels electrons. Using the auxiliary QD excitonic states, we will solve  Eq. \ref{eq:excitonQDCorrection}. By doing this we are including the correction represented by Eq. \ref{eq:Vcorr}. 

%-----------------------  fig QD absorption ------------------------
\begin{figure}[ht]
\centering 
\includegraphics[width=9cm, height=7cm]{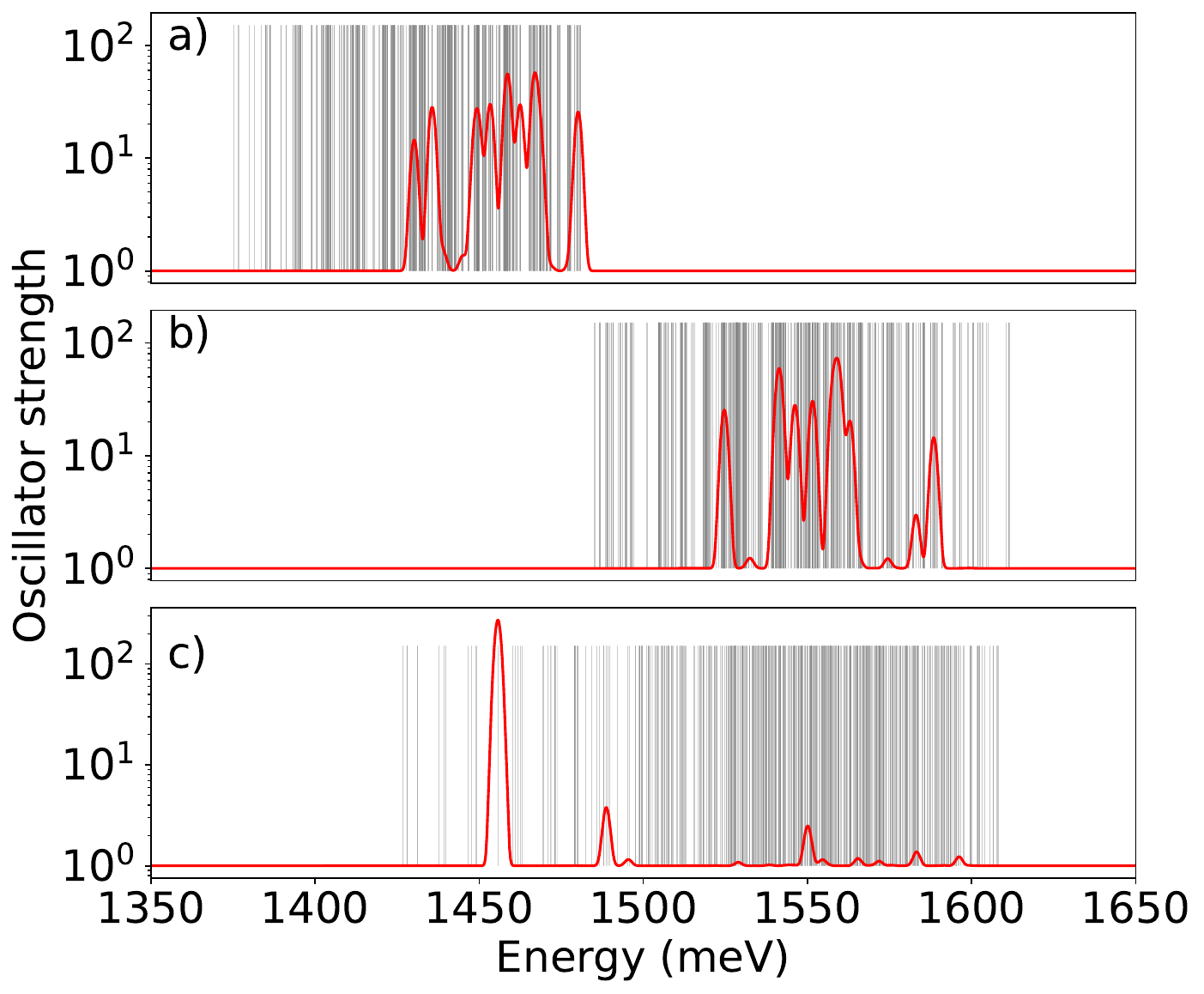}
\caption{
(Color online)  Lowest energy absorption spectrum for a QD with $R_{QD}=20$ nm, $V_0=100$ meV, and $\epsilon = 3.9$ Panel a) shows the single-particle picture, b) includes self-energy and vertex corrections on top of it, and c) also allows for the scattering of configurations. The broadening has been added by hand.} 
\label{fig: QD_absorption}
\end{figure}
%-----------------------------------------

Figure \ref{fig: QD_absorption} shows the absorption spectrum of the QD as a function of the photon frequency. The red line represents the absorption strength for the active transitions, while the vertical black lines correspond to all excitonic energies.

The upper panel shows the non-interacting excitation spectrum. In this situation, there are more peaks than for the auxiliary QD, since the potential correction created mixing between different shells of the auxiliary QD. 

In the center panel, we include the effect of electron-hole interaction, without yet considering scattering between different excitonic configurations. Note that the Coulomb interaction produced a blue shift in the energy in the same way as in the auxiliary QD. 

The lower panel in Figure \ref{fig: QD_absorption} displays the exciton's absorption spectrum, including all terms in the electron-electron interaction matrix elements. Here, we observe a more abrupt redistribution of peak heights, given by correlations between different excitonic configurations. Additionally, the brightest peak experiences a redshift to lower energies after including the correlation between different exciton configurations.

As discussed previously and illustrated schematically in Figure \ref{fig2}, the energy of bound electron-hole pairs can be lower compared to those of localized holes and delocalized electrons, depending on the competition between the electron repelling potential and electron-hole attraction. One can estimate the energy of the type II excitation, in which the hole is inside the QD and the electron is outside,  by considering the energy required to remove one electron from the highest valence band states and place it in the lowest conduction band state, that is delocalized. This transition energy is given by:
\begin{align}
   E_{ex}^\ast = E_g - E_{-1} + \frac{1}{\epsilon}\sum_{k\in VB} V_{-1,k,-1,k}\delta_{\sigma_{-1},\sigma_k}.
\end{align}

Here, $E_g$ represents the material gap, $E_{-1}$ denotes the energy of the highest energy hole state, measured from the top of the valence band, and the last term accounts for the self-energy of this state, which includes interactions with the positive background and exchange interactions with the filled valence band states.

\begin{figure}[ht]
\centering          
\includegraphics[width=9cm, height=7cm]{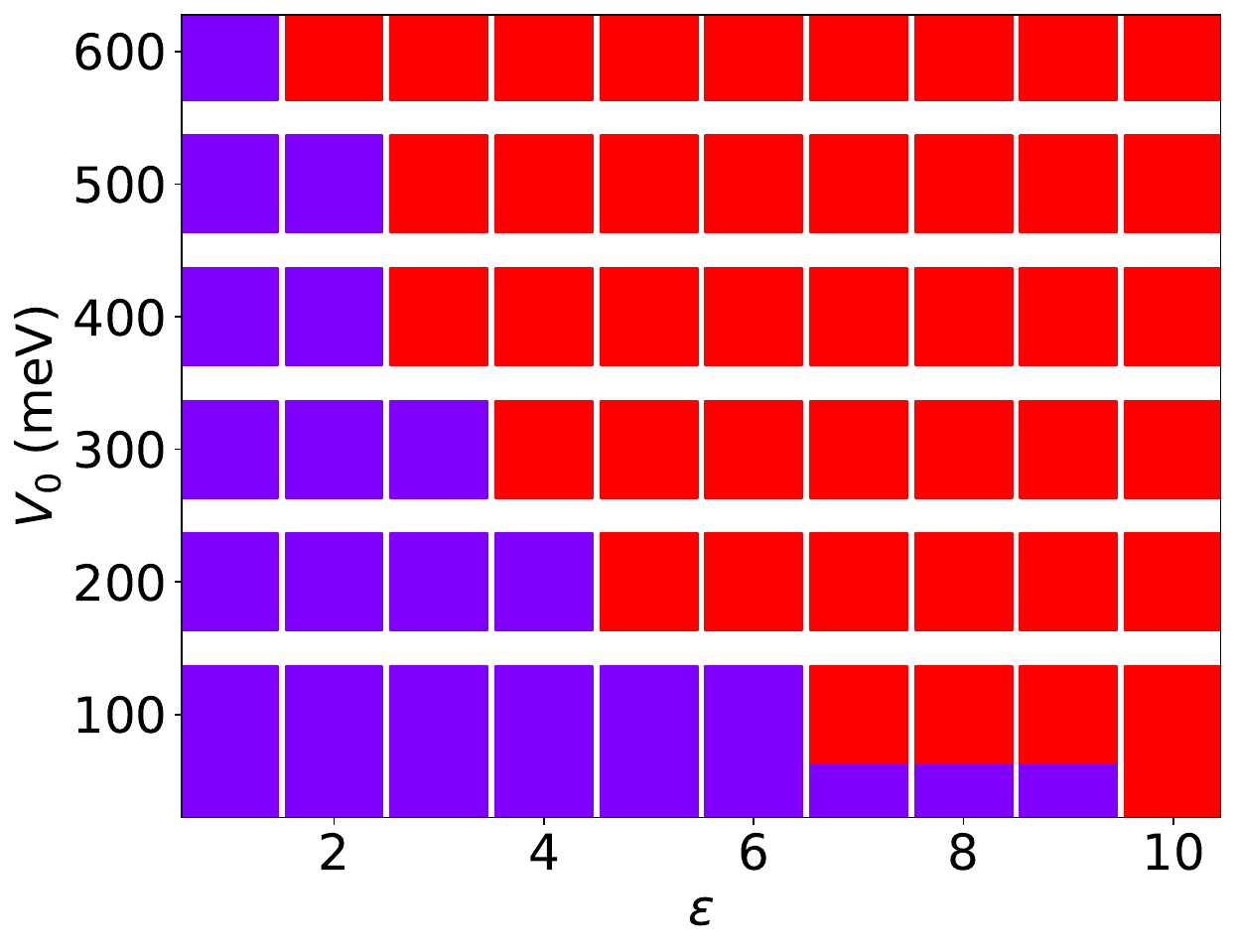}
\caption{
(Color online)  Color map showing in purple the regions in the parameter space $V_0$ vs $\epsilon$ where the transition to a localized electron state has lower energy than the transition to a delocalized one.  } 
\label{fig:phasediagram}
\end{figure}

Figure \ref{fig:phasediagram} shows in the parameter space $V_0$ vs $\epsilon$, the region, represented by purple color, in which the energy of the lowest bright transition is lower than the energy of transition from a localized hole state to a bulk CB state. Note that the optimal situation i.e., exciton is confined, when the interaction is strong (small $\epsilon$) and the confinement is weak (small $V_0$). For $V_0 = 100 $ meV, the situation studied previously, the transition to a localized transition has lower energy than the delocalized one for dielectric constants $\epsilon < 7$.  
A similar qualitative dependency was observed for type II self-assembled QD in reference \cite{IIexcitonsPeeters2001}, in which changing the barrier potential for holes resulted in competition between lowest energy type I and type II excitons obtained using effective mass approximation.

%-----------------------  fig phase diagram ------------------------
%-----------------------------------------

%%%%%%%%%%%%%%%%%%%%%%%%%%%%%%%%%%%%%%%%%%%%%%%%%%%%%%%%%%%%%%%%%%%%%

% Conclusions 
%%%%%%%%%%%%%%%%%%%%%%%%%%%%%%%%%%%%%%%%%%%%%%%%%%%%%%%%%%%%%%%%%%%%%%%%%%%%%%%%%%%%%%%%%%%%%%%
\section{Conclusions}
\label{sect:conclusions}
We have presented a theory of excitons confined in gated $\text{WSe}_\text{2}$ quantum dots. Our work addresses the nontrivial challenge of confining excitons in gated quantum dots, where holes are attracted and the gate potential repels electrons. For exciton to exist the repulsion of electrons by the gate potential must be balanced with the attraction by the Coulomb potential of a hole localized within the dot. Excitons can be successfully confined within the quantum dot through a delicate interplay of these forces, enabling their detection via transport experiments.

We described our WSe2 quantum dot using a computational box containing over a million atoms and employing a tight-binding microscopic Hamiltonian. By applying gates to create a Gaussian confining potential, attractive for holes and repulsive for electrons, we obtained quantum dot energy levels and wavefunctions by expanding the quantum dot wavefunction in bulk band states.

Subsequently, we split our Hamiltonian into an auxiliary Hamiltonian and a perturbation. The auxiliary Hamiltonian confined both electrons and holes. We obtained the excitonic states of the auxiliary quantum dot and expanded the excitonic states of the true quantum dot in this auxiliary basis. In this basis, the repulsive potential of the gate pulled only the electron in the conduction band away from the quantum dot and localized the valence band hole. This approach enabled us to predict a set of parameters for which excitons are trapped in gated quantum dots in 2D materials.
%%%%%%%%%%%%%%%%%%%%%%%%%%%%%%%%%%%%%%%%%%%%%%%%%%%%%%%%%%%%%%%%%%%%%%%%%%%%%%%%%%%%%%%%%%%%%%%%

% Acknowledgments
%%%%%%%%%%%%%%%%%%%%%%%%%%%%%%%%%%%%%%%%%%%%%%%%%%%%%%%%%%%%%%%%%%%%%%%%%%%%%%%%%%%%%%%%%%%%%%%
\begin{acknowledgments}
This work was supported by the  Quantum Sensors Challenge Program QSP-78, High Throughput Networks HTSN-341, and Applied Quantum Computing AQC-04 Challenge Programs at the National Research Council of Canada,  NSERC Discovery Grant No. RGPIN- 2019-05714, NSERC Alliance Quantum Grant No. ALLRP/578466-2022, and University of Ottawa Research Chair in Quantum Theory of Materials, Nanostructures, and Devices. M.B. acknowledges financial support from the National Science Centre, Poland, under Grant No. 2021/43/D/ST3/01989. This research was partly enabled by support provided by the Digital Research Alliance of Canada (alliancecan.ca).
\end{acknowledgments}
%%%%%%%%%%%%%%%%%%%%%%%%%%%%%%%%%%%%%%%%%%%%%%%%%%%%%%%%%%%%%%%%%%%%%%%%%%%%%%%%%%%%%%%%%%%%%%%%

% Bibliography 
%%%%%%%%%%%%%%%%%%%%%%%%%%%%%%%%%%%%%%%%%%%%%%%%%%%%%%%%%%%%%%%%%%%%%%%%%%%%%%%%%%%%%%%%%%%%%%%
\bibliography{main}% Produces the bibliography via BibTeX.

%apsrev4-2.bst 2019-01-14 (MD) hand-edited version of apsrev4-1.bst
%Control: key (0)
%Control: author (8) initials jnrlst
%Control: editor formatted (1) identically to author
%Control: production of article title (0) allowed
%Control: page (0) single
%Control: year (1) truncated
%Control: production of eprint (0) enabled
\providecommand{\noopsort}[1]{}\providecommand{\singleletter}[1]{#1}%
\begin{thebibliography}{56}%
\makeatletter
\providecommand \@ifxundefined [1]{%
 \@ifx{#1\undefined}
}%
\providecommand \@ifnum [1]{%
 \ifnum #1\expandafter \@firstoftwo
 \else \expandafter \@secondoftwo
 \fi
}%
\providecommand \@ifx [1]{%
 \ifx #1\expandafter \@firstoftwo
 \else \expandafter \@secondoftwo
 \fi
}%
\providecommand \natexlab [1]{#1}%
\providecommand \enquote  [1]{``#1''}%
\providecommand \bibnamefont  [1]{#1}%
\providecommand \bibfnamefont [1]{#1}%
\providecommand \citenamefont [1]{#1}%
\providecommand \href@noop [0]{\@secondoftwo}%
\providecommand \href [0]{\begingroup \@sanitize@url \@href}%
\providecommand \@href[1]{\@@startlink{#1}\@@href}%
\providecommand \@@href[1]{\endgroup#1\@@endlink}%
\providecommand \@sanitize@url [0]{\catcode `\\12\catcode `\$12\catcode `\&12\catcode `\#12\catcode `\^12\catcode `\_12\catcode `\%12\relax}%
\providecommand \@@startlink[1]{}%
\providecommand \@@endlink[0]{}%
\providecommand \url  [0]{\begingroup\@sanitize@url \@url }%
\providecommand \@url [1]{\endgroup\@href {#1}{\urlprefix }}%
\providecommand \urlprefix  [0]{URL }%
\providecommand \Eprint [0]{\href }%
\providecommand \doibase [0]{https://doi.org/}%
\providecommand \selectlanguage [0]{\@gobble}%
\providecommand \bibinfo  [0]{\@secondoftwo}%
\providecommand \bibfield  [0]{\@secondoftwo}%
\providecommand \translation [1]{[#1]}%
\providecommand \BibitemOpen [0]{}%
\providecommand \bibitemStop [0]{}%
\providecommand \bibitemNoStop [0]{.\EOS\space}%
\providecommand \EOS [0]{\spacefactor3000\relax}%
\providecommand \BibitemShut  [1]{\csname bibitem#1\endcsname}%
\let\auto@bib@innerbib\@empty
%</preamble>
\bibitem [{\citenamefont {Kadantsev}\ and\ \citenamefont {Hawrylak}(2012)}]{kadantsev2012electronic}%
  \BibitemOpen
  \bibfield  {author} {\bibinfo {author} {\bibfnamefont {E.~S.}\ \bibnamefont {Kadantsev}}\ and\ \bibinfo {author} {\bibfnamefont {P.}~\bibnamefont {Hawrylak}},\ }\bibfield  {title} {\bibinfo {title} {Electronic structure of a single {Mo}{$\text{S}_2$} monolayer},\ }\href {https://doi.org/https://doi.org/10.1016/j.ssc.2012.02.005} {\bibfield  {journal} {\bibinfo  {journal} {Solid state communications}\ }\textbf {\bibinfo {volume} {152}},\ \bibinfo {pages} {909} (\bibinfo {year} {2012})}\BibitemShut {NoStop}%
\bibitem [{\citenamefont {Goh}\ \emph {et~al.}(2020)\citenamefont {Goh}, \citenamefont {Bussolotti}, \citenamefont {Lau}, \citenamefont {Kotekar-Patil}, \citenamefont {Ooi},\ and\ \citenamefont {Chee}}]{Goh2020}%
  \BibitemOpen
  \bibfield  {author} {\bibinfo {author} {\bibfnamefont {K.~E.~J.}\ \bibnamefont {Goh}}, \bibinfo {author} {\bibfnamefont {F.}~\bibnamefont {Bussolotti}}, \bibinfo {author} {\bibfnamefont {C.~S.}\ \bibnamefont {Lau}}, \bibinfo {author} {\bibfnamefont {D.}~\bibnamefont {Kotekar-Patil}}, \bibinfo {author} {\bibfnamefont {Z.~E.}\ \bibnamefont {Ooi}},\ and\ \bibinfo {author} {\bibfnamefont {J.~Y.}\ \bibnamefont {Chee}},\ }\bibfield  {title} {\bibinfo {title} {Toward valley-coupled spin qubits},\ }\href {https://doi.org/https://doi.org/10.1002/qute.201900123} {\bibfield  {journal} {\bibinfo  {journal} {Advanced Quantum Technologies}\ }\textbf {\bibinfo {volume} {3}},\ \bibinfo {pages} {1900123} (\bibinfo {year} {2020})}\BibitemShut {NoStop}%
\bibitem [{\citenamefont {Sierra}\ \emph {et~al.}(2021)\citenamefont {Sierra}, \citenamefont {Fabian}, \citenamefont {Kawakami}, \citenamefont {Roche},\ and\ \citenamefont {Valenzuela}}]{Sierra2021}%
  \BibitemOpen
  \bibfield  {author} {\bibinfo {author} {\bibfnamefont {J.~F.}\ \bibnamefont {Sierra}}, \bibinfo {author} {\bibfnamefont {J.}~\bibnamefont {Fabian}}, \bibinfo {author} {\bibfnamefont {R.~K.}\ \bibnamefont {Kawakami}}, \bibinfo {author} {\bibfnamefont {S.}~\bibnamefont {Roche}},\ and\ \bibinfo {author} {\bibfnamefont {S.~O.}\ \bibnamefont {Valenzuela}},\ }\bibfield  {title} {\bibinfo {title} {Van der waals heterostructures for spintronics and opto-spintronics},\ }\href {https://doi.org/10.1038/s41565-021-00936-x} {\bibfield  {journal} {\bibinfo  {journal} {Nature Nanotechnology}\ }\textbf {\bibinfo {volume} {16}},\ \bibinfo {pages} {856} (\bibinfo {year} {2021})}\BibitemShut {NoStop}%
\bibitem [{\citenamefont {Splendiani}\ \emph {et~al.}(2010)\citenamefont {Splendiani}, \citenamefont {Sun}, \citenamefont {Zhang}, \citenamefont {Li}, \citenamefont {Kim}, \citenamefont {Chim}, \citenamefont {Galli},\ and\ \citenamefont {Wang}}]{splendiani2010emerging}%
  \BibitemOpen
  \bibfield  {author} {\bibinfo {author} {\bibfnamefont {A.}~\bibnamefont {Splendiani}}, \bibinfo {author} {\bibfnamefont {L.}~\bibnamefont {Sun}}, \bibinfo {author} {\bibfnamefont {Y.}~\bibnamefont {Zhang}}, \bibinfo {author} {\bibfnamefont {T.}~\bibnamefont {Li}}, \bibinfo {author} {\bibfnamefont {J.}~\bibnamefont {Kim}}, \bibinfo {author} {\bibfnamefont {C.-Y.}\ \bibnamefont {Chim}}, \bibinfo {author} {\bibfnamefont {G.}~\bibnamefont {Galli}},\ and\ \bibinfo {author} {\bibfnamefont {F.}~\bibnamefont {Wang}},\ }\bibfield  {title} {\bibinfo {title} {Emerging photoluminescence in monolayer {Mo}{$\text{S}_2$}},\ }\href {https://doi.org/10.1021/nl903868w} {\bibfield  {journal} {\bibinfo  {journal} {Nano letters}\ }\textbf {\bibinfo {volume} {10}},\ \bibinfo {pages} {1271} (\bibinfo {year} {2010})}\BibitemShut {NoStop}%
\bibitem [{\citenamefont {Roch}\ \emph {et~al.}(2019)\citenamefont {Roch}, \citenamefont {Froehlicher}, \citenamefont {Leisgang}, \citenamefont {Makk}, \citenamefont {Watanabe}, \citenamefont {Taniguchi},\ and\ \citenamefont {Warburton}}]{roch2019spin}%
  \BibitemOpen
  \bibfield  {author} {\bibinfo {author} {\bibfnamefont {J.~G.}\ \bibnamefont {Roch}}, \bibinfo {author} {\bibfnamefont {G.}~\bibnamefont {Froehlicher}}, \bibinfo {author} {\bibfnamefont {N.}~\bibnamefont {Leisgang}}, \bibinfo {author} {\bibfnamefont {P.}~\bibnamefont {Makk}}, \bibinfo {author} {\bibfnamefont {K.}~\bibnamefont {Watanabe}}, \bibinfo {author} {\bibfnamefont {T.}~\bibnamefont {Taniguchi}},\ and\ \bibinfo {author} {\bibfnamefont {R.~J.}\ \bibnamefont {Warburton}},\ }\bibfield  {title} {\bibinfo {title} {Spin-polarized electrons in monolayer {Mo}{$\text{S}_2$}},\ }\href {https://doi.org/10.1038/s41565-019-0397-y} {\bibfield  {journal} {\bibinfo  {journal} {Nature nanotechnology}\ }\textbf {\bibinfo {volume} {14}},\ \bibinfo {pages} {432} (\bibinfo {year} {2019})}\BibitemShut {NoStop}%
\bibitem [{\citenamefont {Scrace}\ \emph {et~al.}(2015)\citenamefont {Scrace}, \citenamefont {Tsai}, \citenamefont {Barman}, \citenamefont {Schweidenback}, \citenamefont {Petrou}, \citenamefont {Kioseoglou}, \citenamefont {Ozfidan}, \citenamefont {Korkusinski},\ and\ \citenamefont {Hawrylak}}]{scrace2015magnetoluminescence}%
  \BibitemOpen
  \bibfield  {author} {\bibinfo {author} {\bibfnamefont {T.}~\bibnamefont {Scrace}}, \bibinfo {author} {\bibfnamefont {Y.}~\bibnamefont {Tsai}}, \bibinfo {author} {\bibfnamefont {B.}~\bibnamefont {Barman}}, \bibinfo {author} {\bibfnamefont {L.}~\bibnamefont {Schweidenback}}, \bibinfo {author} {\bibfnamefont {A.}~\bibnamefont {Petrou}}, \bibinfo {author} {\bibfnamefont {G.}~\bibnamefont {Kioseoglou}}, \bibinfo {author} {\bibfnamefont {I.}~\bibnamefont {Ozfidan}}, \bibinfo {author} {\bibfnamefont {M.}~\bibnamefont {Korkusinski}},\ and\ \bibinfo {author} {\bibfnamefont {P.}~\bibnamefont {Hawrylak}},\ }\bibfield  {title} {\bibinfo {title} {Magnetoluminescence and valley polarized state of a two-dimensional electron gas in {W}{$\text{S}_2$} monolayers},\ }\href {https://doi.org/10.1038/nnano.2015.78} {\bibfield  {journal} {\bibinfo  {journal} {Nature nanotechnology}\ }\textbf {\bibinfo {volume} {10}},\ \bibinfo {pages} {603} (\bibinfo {year} {2015})}\BibitemShut {NoStop}%
\bibitem [{\citenamefont {Van~Tuan}\ \emph {et~al.}(2019)\citenamefont {Van~Tuan}, \citenamefont {Jones}, \citenamefont {Yang}, \citenamefont {Xu},\ and\ \citenamefont {Dery}}]{van2019virtual}%
  \BibitemOpen
  \bibfield  {author} {\bibinfo {author} {\bibfnamefont {D.}~\bibnamefont {Van~Tuan}}, \bibinfo {author} {\bibfnamefont {A.~M.}\ \bibnamefont {Jones}}, \bibinfo {author} {\bibfnamefont {M.}~\bibnamefont {Yang}}, \bibinfo {author} {\bibfnamefont {X.}~\bibnamefont {Xu}},\ and\ \bibinfo {author} {\bibfnamefont {H.}~\bibnamefont {Dery}},\ }\bibfield  {title} {\bibinfo {title} {Virtual trions in the photoluminescence of monolayer transition-metal dichalcogenides},\ }\href {https://doi.org/10.1103/PhysRevLett.122.217401} {\bibfield  {journal} {\bibinfo  {journal} {Phys. Rev. Lett.}\ }\textbf {\bibinfo {volume} {122}},\ \bibinfo {pages} {217401} (\bibinfo {year} {2019})}\BibitemShut {NoStop}%
\bibitem [{\citenamefont {Komsa}\ and\ \citenamefont {Krasheninnikov}(2012)}]{komsa2012two}%
  \BibitemOpen
  \bibfield  {author} {\bibinfo {author} {\bibfnamefont {H.-P.}\ \bibnamefont {Komsa}}\ and\ \bibinfo {author} {\bibfnamefont {A.~V.}\ \bibnamefont {Krasheninnikov}},\ }\bibfield  {title} {\bibinfo {title} {Two-dimensional transition metal dichalcogenide alloys: stability and electronic properties},\ }\href {https://doi.org/10.1021/jz301673x} {\bibfield  {journal} {\bibinfo  {journal} {The journal of physical chemistry letters}\ }\textbf {\bibinfo {volume} {3}},\ \bibinfo {pages} {3652} (\bibinfo {year} {2012})}\BibitemShut {NoStop}%
\bibitem [{\citenamefont {Boddison-Chouinard}\ \emph {et~al.}(2021)\citenamefont {Boddison-Chouinard}, \citenamefont {Bogan}, \citenamefont {Fong}, \citenamefont {Watanabe}, \citenamefont {Taniguchi}, \citenamefont {Studenikin}, \citenamefont {Sachrajda}, \citenamefont {Korkusinski}, \citenamefont {Altintas}, \citenamefont {Bieniek}, \citenamefont {Hawrylak}, \citenamefont {Luican-Mayer},\ and\ \citenamefont {Gaudreau}}]{boddison2021gate}%
  \BibitemOpen
  \bibfield  {author} {\bibinfo {author} {\bibfnamefont {J.}~\bibnamefont {Boddison-Chouinard}}, \bibinfo {author} {\bibfnamefont {A.}~\bibnamefont {Bogan}}, \bibinfo {author} {\bibfnamefont {N.}~\bibnamefont {Fong}}, \bibinfo {author} {\bibfnamefont {K.}~\bibnamefont {Watanabe}}, \bibinfo {author} {\bibfnamefont {T.}~\bibnamefont {Taniguchi}}, \bibinfo {author} {\bibfnamefont {S.}~\bibnamefont {Studenikin}}, \bibinfo {author} {\bibfnamefont {A.}~\bibnamefont {Sachrajda}}, \bibinfo {author} {\bibfnamefont {M.}~\bibnamefont {Korkusinski}}, \bibinfo {author} {\bibfnamefont {A.}~\bibnamefont {Altintas}}, \bibinfo {author} {\bibfnamefont {M.}~\bibnamefont {Bieniek}}, \bibinfo {author} {\bibfnamefont {P.}~\bibnamefont {Hawrylak}}, \bibinfo {author} {\bibfnamefont {A.}~\bibnamefont {Luican-Mayer}},\ and\ \bibinfo {author} {\bibfnamefont {L.}~\bibnamefont {Gaudreau}},\ }\bibfield  {title} {\bibinfo {title} {{Gate-controlled quantum dots in monolayer WSe2}},\ }\href {https://doi.org/10.1063/5.0062838} {\bibfield
  {journal} {\bibinfo  {journal} {Applied Physics Letters}\ }\textbf {\bibinfo {volume} {119}},\ \bibinfo {pages} {133104} (\bibinfo {year} {2021})}\BibitemShut {NoStop}%
\bibitem [{\citenamefont {Boddison-Chouinard}\ \emph {et~al.}(2023)\citenamefont {Boddison-Chouinard}, \citenamefont {Bogan}, \citenamefont {Barrios}, \citenamefont {Lapointe}, \citenamefont {Watanabe}, \citenamefont {Taniguchi}, \citenamefont {Paw{\l}owski}, \citenamefont {Miravet}, \citenamefont {Bieniek}, \citenamefont {Hawrylak}, \citenamefont {Luican-Mayer},\ and\ \citenamefont {Gaudreau}}]{Boddison20231DChannel}%
  \BibitemOpen
  \bibfield  {author} {\bibinfo {author} {\bibfnamefont {J.}~\bibnamefont {Boddison-Chouinard}}, \bibinfo {author} {\bibfnamefont {A.}~\bibnamefont {Bogan}}, \bibinfo {author} {\bibfnamefont {P.}~\bibnamefont {Barrios}}, \bibinfo {author} {\bibfnamefont {J.}~\bibnamefont {Lapointe}}, \bibinfo {author} {\bibfnamefont {K.}~\bibnamefont {Watanabe}}, \bibinfo {author} {\bibfnamefont {T.}~\bibnamefont {Taniguchi}}, \bibinfo {author} {\bibfnamefont {J.}~\bibnamefont {Paw{\l}owski}}, \bibinfo {author} {\bibfnamefont {D.}~\bibnamefont {Miravet}}, \bibinfo {author} {\bibfnamefont {M.}~\bibnamefont {Bieniek}}, \bibinfo {author} {\bibfnamefont {P.}~\bibnamefont {Hawrylak}}, \bibinfo {author} {\bibfnamefont {A.}~\bibnamefont {Luican-Mayer}},\ and\ \bibinfo {author} {\bibfnamefont {L.}~\bibnamefont {Gaudreau}},\ }\bibfield  {title} {\bibinfo {title} {Anomalous conductance quantization of a one-dimensional channel in monolayer {W}{$\text{Se}_2$}},\ }\href {https://doi.org/10.1038/s41699-023-00407-y} {\bibfield  {journal}
  {\bibinfo  {journal} {npj 2D Materials and Applications}\ }\textbf {\bibinfo {volume} {7}},\ \bibinfo {pages} {50} (\bibinfo {year} {2023})}\BibitemShut {NoStop}%
\bibitem [{\citenamefont {Mueller}\ and\ \citenamefont {Malic}(2018)}]{mueller2018exciton}%
  \BibitemOpen
  \bibfield  {author} {\bibinfo {author} {\bibfnamefont {T.}~\bibnamefont {Mueller}}\ and\ \bibinfo {author} {\bibfnamefont {E.}~\bibnamefont {Malic}},\ }\bibfield  {title} {\bibinfo {title} {Exciton physics and device application of two-dimensional transition metal dichalcogenide semiconductors},\ }\href {https://doi.org/10.1038/s41699-018-0074-2} {\bibfield  {journal} {\bibinfo  {journal} {npj 2D Materials and Applications}\ }\textbf {\bibinfo {volume} {2}},\ \bibinfo {pages} {1} (\bibinfo {year} {2018})}\BibitemShut {NoStop}%
\bibitem [{\citenamefont {Schneider}\ \emph {et~al.}(2018)\citenamefont {Schneider}, \citenamefont {Glazov}, \citenamefont {Korn}, \citenamefont {H{\"o}fling},\ and\ \citenamefont {Urbaszek}}]{schneider2018two}%
  \BibitemOpen
  \bibfield  {author} {\bibinfo {author} {\bibfnamefont {C.}~\bibnamefont {Schneider}}, \bibinfo {author} {\bibfnamefont {M.~M.}\ \bibnamefont {Glazov}}, \bibinfo {author} {\bibfnamefont {T.}~\bibnamefont {Korn}}, \bibinfo {author} {\bibfnamefont {S.}~\bibnamefont {H{\"o}fling}},\ and\ \bibinfo {author} {\bibfnamefont {B.}~\bibnamefont {Urbaszek}},\ }\bibfield  {title} {\bibinfo {title} {Two-dimensional semiconductors in the regime of strong light-matter coupling},\ }\href {https://doi.org/10.1038/s41467-018-04866-6} {\bibfield  {journal} {\bibinfo  {journal} {Nature communications}\ }\textbf {\bibinfo {volume} {9}},\ \bibinfo {pages} {1} (\bibinfo {year} {2018})}\BibitemShut {NoStop}%
\bibitem [{\citenamefont {Conley}\ \emph {et~al.}(2013)\citenamefont {Conley}, \citenamefont {Wang}, \citenamefont {Ziegler}, \citenamefont {Haglund~Jr}, \citenamefont {Pantelides},\ and\ \citenamefont {Bolotin}}]{conley2013bandgap}%
  \BibitemOpen
  \bibfield  {author} {\bibinfo {author} {\bibfnamefont {H.~J.}\ \bibnamefont {Conley}}, \bibinfo {author} {\bibfnamefont {B.}~\bibnamefont {Wang}}, \bibinfo {author} {\bibfnamefont {J.~I.}\ \bibnamefont {Ziegler}}, \bibinfo {author} {\bibfnamefont {R.~F.}\ \bibnamefont {Haglund~Jr}}, \bibinfo {author} {\bibfnamefont {S.~T.}\ \bibnamefont {Pantelides}},\ and\ \bibinfo {author} {\bibfnamefont {K.~I.}\ \bibnamefont {Bolotin}},\ }\bibfield  {title} {\bibinfo {title} {Bandgap engineering of strained monolayer and bilayer {Mo}{$\text{S}_2$}},\ }\href {https://doi.org/10.1021/nl4014748} {\bibfield  {journal} {\bibinfo  {journal} {Nano letters}\ }\textbf {\bibinfo {volume} {13}},\ \bibinfo {pages} {3626} (\bibinfo {year} {2013})}\BibitemShut {NoStop}%
\bibitem [{\citenamefont {Manzeli}\ \emph {et~al.}(2017)\citenamefont {Manzeli}, \citenamefont {Ovchinnikov}, \citenamefont {Pasquier}, \citenamefont {Yazyev},\ and\ \citenamefont {Kis}}]{Manzeli_Kis_2017}%
  \BibitemOpen
  \bibfield  {author} {\bibinfo {author} {\bibfnamefont {S.}~\bibnamefont {Manzeli}}, \bibinfo {author} {\bibfnamefont {D.}~\bibnamefont {Ovchinnikov}}, \bibinfo {author} {\bibfnamefont {D.}~\bibnamefont {Pasquier}}, \bibinfo {author} {\bibfnamefont {O.~V.}\ \bibnamefont {Yazyev}},\ and\ \bibinfo {author} {\bibfnamefont {A.}~\bibnamefont {Kis}},\ }\bibfield  {title} {\bibinfo {title} {2d transition metal dichalcogenides},\ }\href {https://doi.org/10.1038/natrevmats.2017.33} {\bibfield  {journal} {\bibinfo  {journal} {Nature Reviews Materials}\ }\textbf {\bibinfo {volume} {2}},\ \bibinfo {pages} {17033} (\bibinfo {year} {2017})}\BibitemShut {NoStop}%
\bibitem [{\citenamefont {Castro~Neto}\ \emph {et~al.}(2009)\citenamefont {Castro~Neto}, \citenamefont {Guinea}, \citenamefont {Peres}, \citenamefont {Novoselov},\ and\ \citenamefont {Geim}}]{CastroNeto_Geim_2009}%
  \BibitemOpen
  \bibfield  {author} {\bibinfo {author} {\bibfnamefont {A.~H.}\ \bibnamefont {Castro~Neto}}, \bibinfo {author} {\bibfnamefont {F.}~\bibnamefont {Guinea}}, \bibinfo {author} {\bibfnamefont {N.~M.~R.}\ \bibnamefont {Peres}}, \bibinfo {author} {\bibfnamefont {K.~S.}\ \bibnamefont {Novoselov}},\ and\ \bibinfo {author} {\bibfnamefont {A.~K.}\ \bibnamefont {Geim}},\ }\bibfield  {title} {\bibinfo {title} {The electronic properties of graphene},\ }\href {https://doi.org/10.1103/RevModPhys.81.109} {\bibfield  {journal} {\bibinfo  {journal} {Rev. Mod. Phys.}\ }\textbf {\bibinfo {volume} {81}},\ \bibinfo {pages} {109} (\bibinfo {year} {2009})}\BibitemShut {NoStop}%
\bibitem [{\citenamefont {Mak}\ \emph {et~al.}(2010)\citenamefont {Mak}, \citenamefont {Lee}, \citenamefont {Hone}, \citenamefont {Shan},\ and\ \citenamefont {Heinz}}]{Mak_Heinz_2010}%
  \BibitemOpen
  \bibfield  {author} {\bibinfo {author} {\bibfnamefont {K.~F.}\ \bibnamefont {Mak}}, \bibinfo {author} {\bibfnamefont {C.}~\bibnamefont {Lee}}, \bibinfo {author} {\bibfnamefont {J.}~\bibnamefont {Hone}}, \bibinfo {author} {\bibfnamefont {J.}~\bibnamefont {Shan}},\ and\ \bibinfo {author} {\bibfnamefont {T.~F.}\ \bibnamefont {Heinz}},\ }\bibfield  {title} {\bibinfo {title} {Atomically thin {Mo}{$\text{S}_2$}: A new direct-gap semiconductor},\ }\href {https://doi.org/10.1103/PhysRevLett.105.136805} {\bibfield  {journal} {\bibinfo  {journal} {Phys. Rev. Lett.}\ }\textbf {\bibinfo {volume} {105}},\ \bibinfo {pages} {136805} (\bibinfo {year} {2010})}\BibitemShut {NoStop}%
\bibitem [{\citenamefont {Alt{\i}nta{\c{s}}}\ \emph {et~al.}(2021)\citenamefont {Alt{\i}nta{\c{s}}}, \citenamefont {Bieniek}, \citenamefont {Dusko}, \citenamefont {Korkusi{\'n}ski}, \citenamefont {Paw{\l}owski},\ and\ \citenamefont {Hawrylak}}]{altintacs2021spin}%
  \BibitemOpen
  \bibfield  {author} {\bibinfo {author} {\bibfnamefont {A.}~\bibnamefont {Alt{\i}nta{\c{s}}}}, \bibinfo {author} {\bibfnamefont {M.}~\bibnamefont {Bieniek}}, \bibinfo {author} {\bibfnamefont {A.}~\bibnamefont {Dusko}}, \bibinfo {author} {\bibfnamefont {M.}~\bibnamefont {Korkusi{\'n}ski}}, \bibinfo {author} {\bibfnamefont {J.}~\bibnamefont {Paw{\l}owski}},\ and\ \bibinfo {author} {\bibfnamefont {P.}~\bibnamefont {Hawrylak}},\ }\bibfield  {title} {\bibinfo {title} {Spin-valley qubits in gated quantum dots in a single layer of transition metal dichalcogenides},\ }\href {https://doi.org/10.1103/PhysRevB.104.195412} {\bibfield  {journal} {\bibinfo  {journal} {Physical Review B}\ }\textbf {\bibinfo {volume} {104}},\ \bibinfo {pages} {195412} (\bibinfo {year} {2021})}\BibitemShut {NoStop}%
\bibitem [{\citenamefont {Bieniek}\ \emph {et~al.}(2020)\citenamefont {Bieniek}, \citenamefont {Szulakowska},\ and\ \citenamefont {Hawrylak}}]{bieniek2020effect}%
  \BibitemOpen
  \bibfield  {author} {\bibinfo {author} {\bibfnamefont {M.}~\bibnamefont {Bieniek}}, \bibinfo {author} {\bibfnamefont {L.}~\bibnamefont {Szulakowska}},\ and\ \bibinfo {author} {\bibfnamefont {P.}~\bibnamefont {Hawrylak}},\ }\bibfield  {title} {\bibinfo {title} {Effect of valley, spin, and band nesting on the electronic properties of gated quantum dots in a single layer of transition metal dichalcogenides},\ }\href {https://doi.org/10.1103/PhysRevB.101.035401} {\bibfield  {journal} {\bibinfo  {journal} {Physical Review B}\ }\textbf {\bibinfo {volume} {101}},\ \bibinfo {pages} {035401} (\bibinfo {year} {2020})}\BibitemShut {NoStop}%
\bibitem [{\citenamefont {Bieniek}\ \emph {et~al.}(2018)\citenamefont {Bieniek}, \citenamefont {Korkusi{\'n}ski}, \citenamefont {Szulakowska}, \citenamefont {Potasz}, \citenamefont {Ozfidan},\ and\ \citenamefont {Hawrylak}}]{bieniek2018band}%
  \BibitemOpen
  \bibfield  {author} {\bibinfo {author} {\bibfnamefont {M.}~\bibnamefont {Bieniek}}, \bibinfo {author} {\bibfnamefont {M.}~\bibnamefont {Korkusi{\'n}ski}}, \bibinfo {author} {\bibfnamefont {L.}~\bibnamefont {Szulakowska}}, \bibinfo {author} {\bibfnamefont {P.}~\bibnamefont {Potasz}}, \bibinfo {author} {\bibfnamefont {I.}~\bibnamefont {Ozfidan}},\ and\ \bibinfo {author} {\bibfnamefont {P.}~\bibnamefont {Hawrylak}},\ }\bibfield  {title} {\bibinfo {title} {Band nesting, massive dirac fermions, and valley land{\'e} and zeeman effects in transition metal dichalcogenides: A tight-binding model},\ }\href {https://doi.org/10.1103/PhysRevB.97.085153} {\bibfield  {journal} {\bibinfo  {journal} {Physical Review B}\ }\textbf {\bibinfo {volume} {97}},\ \bibinfo {pages} {085153} (\bibinfo {year} {2018})}\BibitemShut {NoStop}%
\bibitem [{\citenamefont {Miravet}\ \emph {et~al.}(2023)\citenamefont {Miravet}, \citenamefont {Alt\ifmmode \imath \else \i \fi{}nta\ifmmode~\mbox{\c{s}}\else \c{s}\fi{}}, \citenamefont {Rodrigues}, \citenamefont {Bieniek}, \citenamefont {Korkusinski},\ and\ \citenamefont {Hawrylak}}]{MiravetPRB2023holesQD}%
  \BibitemOpen
  \bibfield  {author} {\bibinfo {author} {\bibfnamefont {D.}~\bibnamefont {Miravet}}, \bibinfo {author} {\bibfnamefont {A.}~\bibnamefont {Alt\ifmmode \imath \else \i \fi{}nta\ifmmode~\mbox{\c{s}}\else \c{s}\fi{}}}, \bibinfo {author} {\bibfnamefont {A.~W.}\ \bibnamefont {Rodrigues}}, \bibinfo {author} {\bibfnamefont {M.}~\bibnamefont {Bieniek}}, \bibinfo {author} {\bibfnamefont {M.}~\bibnamefont {Korkusinski}},\ and\ \bibinfo {author} {\bibfnamefont {P.}~\bibnamefont {Hawrylak}},\ }\bibfield  {title} {\bibinfo {title} {Interacting holes in gated ${\mathrm{wse}}_{2}$ quantum dots},\ }\href {https://doi.org/10.1103/PhysRevB.108.195407} {\bibfield  {journal} {\bibinfo  {journal} {Phys. Rev. B}\ }\textbf {\bibinfo {volume} {108}},\ \bibinfo {pages} {195407} (\bibinfo {year} {2023})}\BibitemShut {NoStop}%
\bibitem [{\citenamefont {Otsuka}\ \emph {et~al.}(2016)\citenamefont {Otsuka}, \citenamefont {Yunoki},\ and\ \citenamefont {Sorella}}]{OtsukaCriticalityMott2016}%
  \BibitemOpen
  \bibfield  {author} {\bibinfo {author} {\bibfnamefont {Y.}~\bibnamefont {Otsuka}}, \bibinfo {author} {\bibfnamefont {S.}~\bibnamefont {Yunoki}},\ and\ \bibinfo {author} {\bibfnamefont {S.}~\bibnamefont {Sorella}},\ }\bibfield  {title} {\bibinfo {title} {Universal quantum criticality in the metal-insulator transition of two-dimensional interacting {D}irac electrons},\ }\href {https://doi.org/10.1103/PhysRevX.6.011029} {\bibfield  {journal} {\bibinfo  {journal} {Phys. Rev. X}\ }\textbf {\bibinfo {volume} {6}},\ \bibinfo {pages} {011029} (\bibinfo {year} {2016})}\BibitemShut {NoStop}%
\bibitem [{\citenamefont {Bieniek}\ \emph {et~al.}(2022)\citenamefont {Bieniek}, \citenamefont {Sadecka}, \citenamefont {Szulakowska},\ and\ \citenamefont {Hawrylak}}]{BieniekExcitons2022}%
  \BibitemOpen
  \bibfield  {author} {\bibinfo {author} {\bibfnamefont {M.}~\bibnamefont {Bieniek}}, \bibinfo {author} {\bibfnamefont {K.}~\bibnamefont {Sadecka}}, \bibinfo {author} {\bibfnamefont {L.}~\bibnamefont {Szulakowska}},\ and\ \bibinfo {author} {\bibfnamefont {P.}~\bibnamefont {Hawrylak}},\ }\bibfield  {title} {\bibinfo {title} {Theory of excitons in atomically thin semiconductors: Tight-binding approach},\ }\bibfield  {journal} {\bibinfo  {journal} {Nanomaterials}\ }\textbf {\bibinfo {volume} {12}},\ \href {https://doi.org/10.3390/nano12091582} {10.3390/nano12091582} (\bibinfo {year} {2022})\BibitemShut {NoStop}%
\bibitem [{\citenamefont {Borges}\ \emph {et~al.}(2023)\citenamefont {Borges}, \citenamefont {J\'unior}, \citenamefont {Brand\~ao}, \citenamefont {Liu}, \citenamefont {Pereira}, \citenamefont {Xie}, \citenamefont {Qu},\ and\ \citenamefont {Alcalde}}]{BorgesexcitonQ2023}%
  \BibitemOpen
  \bibfield  {author} {\bibinfo {author} {\bibfnamefont {H.~S.}\ \bibnamefont {Borges}}, \bibinfo {author} {\bibfnamefont {C.~A.~N.}\ \bibnamefont {J\'unior}}, \bibinfo {author} {\bibfnamefont {D.~S.}\ \bibnamefont {Brand\~ao}}, \bibinfo {author} {\bibfnamefont {F.}~\bibnamefont {Liu}}, \bibinfo {author} {\bibfnamefont {V.~V.~R.}\ \bibnamefont {Pereira}}, \bibinfo {author} {\bibfnamefont {S.~J.}\ \bibnamefont {Xie}}, \bibinfo {author} {\bibfnamefont {F.}~\bibnamefont {Qu}},\ and\ \bibinfo {author} {\bibfnamefont {A.~M.}\ \bibnamefont {Alcalde}},\ }\bibfield  {title} {\bibinfo {title} {Persistent entanglement of valley exciton qubits in transition metal dichalcogenides integrated into a bimodal optical cavity},\ }\href {https://doi.org/10.1103/PhysRevB.107.035404} {\bibfield  {journal} {\bibinfo  {journal} {Phys. Rev. B}\ }\textbf {\bibinfo {volume} {107}},\ \bibinfo {pages} {035404} (\bibinfo {year} {2023})}\BibitemShut {NoStop}%
\bibitem [{\citenamefont {Wu}\ \emph {et~al.}(2015)\citenamefont {Wu}, \citenamefont {Qu},\ and\ \citenamefont {MacDonald}}]{WuExcitonMoS2_2015}%
  \BibitemOpen
  \bibfield  {author} {\bibinfo {author} {\bibfnamefont {F.}~\bibnamefont {Wu}}, \bibinfo {author} {\bibfnamefont {F.}~\bibnamefont {Qu}},\ and\ \bibinfo {author} {\bibfnamefont {A.~H.}\ \bibnamefont {MacDonald}},\ }\bibfield  {title} {\bibinfo {title} {Exciton band structure of monolayer {Mo}{$\text{S}_2$}},\ }\href {https://doi.org/10.1103/PhysRevB.91.075310} {\bibfield  {journal} {\bibinfo  {journal} {Phys. Rev. B}\ }\textbf {\bibinfo {volume} {91}},\ \bibinfo {pages} {075310} (\bibinfo {year} {2015})}\BibitemShut {NoStop}%
\bibitem [{\citenamefont {Goryca}\ \emph {et~al.}(2019)\citenamefont {Goryca}, \citenamefont {Li}, \citenamefont {Stier}, \citenamefont {Taniguchi}, \citenamefont {Watanabe}, \citenamefont {Courtade}, \citenamefont {Shree}, \citenamefont {Robert}, \citenamefont {Urbaszek}, \citenamefont {Marie},\ and\ \citenamefont {Crooker}}]{Goryca2019}%
  \BibitemOpen
  \bibfield  {author} {\bibinfo {author} {\bibfnamefont {M.}~\bibnamefont {Goryca}}, \bibinfo {author} {\bibfnamefont {J.}~\bibnamefont {Li}}, \bibinfo {author} {\bibfnamefont {A.~V.}\ \bibnamefont {Stier}}, \bibinfo {author} {\bibfnamefont {T.}~\bibnamefont {Taniguchi}}, \bibinfo {author} {\bibfnamefont {K.}~\bibnamefont {Watanabe}}, \bibinfo {author} {\bibfnamefont {E.}~\bibnamefont {Courtade}}, \bibinfo {author} {\bibfnamefont {S.}~\bibnamefont {Shree}}, \bibinfo {author} {\bibfnamefont {C.}~\bibnamefont {Robert}}, \bibinfo {author} {\bibfnamefont {B.}~\bibnamefont {Urbaszek}}, \bibinfo {author} {\bibfnamefont {X.}~\bibnamefont {Marie}},\ and\ \bibinfo {author} {\bibfnamefont {S.~A.}\ \bibnamefont {Crooker}},\ }\bibfield  {title} {\bibinfo {title} {Revealing exciton masses and dielectric properties of monolayer semiconductors with high magnetic fields},\ }\href {https://doi.org/10.1038/s41467-019-12180-y} {\bibfield  {journal} {\bibinfo  {journal} {Nature Communications}\ }\textbf {\bibinfo {volume}
  {10}},\ \bibinfo {pages} {4172} (\bibinfo {year} {2019})}\BibitemShut {NoStop}%
\bibitem [{\citenamefont {Thureja}\ \emph {et~al.}(2022)\citenamefont {Thureja}, \citenamefont {Imamoglu}, \citenamefont {Smole{\'{n}}ski}, \citenamefont {Amelio}, \citenamefont {Popert}, \citenamefont {Chervy}, \citenamefont {Lu}, \citenamefont {Liu}, \citenamefont {Barmak}, \citenamefont {Watanabe}, \citenamefont {Taniguchi}, \citenamefont {Norris}, \citenamefont {Kroner},\ and\ \citenamefont {Murthy}}]{Thureja2022}%
  \BibitemOpen
  \bibfield  {author} {\bibinfo {author} {\bibfnamefont {D.}~\bibnamefont {Thureja}}, \bibinfo {author} {\bibfnamefont {A.}~\bibnamefont {Imamoglu}}, \bibinfo {author} {\bibfnamefont {T.}~\bibnamefont {Smole{\'{n}}ski}}, \bibinfo {author} {\bibfnamefont {I.}~\bibnamefont {Amelio}}, \bibinfo {author} {\bibfnamefont {A.}~\bibnamefont {Popert}}, \bibinfo {author} {\bibfnamefont {T.}~\bibnamefont {Chervy}}, \bibinfo {author} {\bibfnamefont {X.}~\bibnamefont {Lu}}, \bibinfo {author} {\bibfnamefont {S.}~\bibnamefont {Liu}}, \bibinfo {author} {\bibfnamefont {K.}~\bibnamefont {Barmak}}, \bibinfo {author} {\bibfnamefont {K.}~\bibnamefont {Watanabe}}, \bibinfo {author} {\bibfnamefont {T.}~\bibnamefont {Taniguchi}}, \bibinfo {author} {\bibfnamefont {D.~J.}\ \bibnamefont {Norris}}, \bibinfo {author} {\bibfnamefont {M.}~\bibnamefont {Kroner}},\ and\ \bibinfo {author} {\bibfnamefont {P.~A.}\ \bibnamefont {Murthy}},\ }\bibfield  {title} {\bibinfo {title} {Electrically tunable quantum confinement of neutral excitons},\
  }\href {https://doi.org/10.1038/s41586-022-04634-z} {\bibfield  {journal} {\bibinfo  {journal} {Nature}\ }\textbf {\bibinfo {volume} {606}},\ \bibinfo {pages} {298} (\bibinfo {year} {2022})}\BibitemShut {NoStop}%
\bibitem [{\citenamefont {Le}\ \emph {et~al.}(2015)\citenamefont {Le}, \citenamefont {Barinov}, \citenamefont {Preciado}, \citenamefont {Isarraraz}, \citenamefont {Tanabe}, \citenamefont {Komesu}, \citenamefont {Troha}, \citenamefont {Bartels}, \citenamefont {Rahman},\ and\ \citenamefont {Dowben}}]{le2015spin}%
  \BibitemOpen
  \bibfield  {author} {\bibinfo {author} {\bibfnamefont {D.}~\bibnamefont {Le}}, \bibinfo {author} {\bibfnamefont {A.}~\bibnamefont {Barinov}}, \bibinfo {author} {\bibfnamefont {E.}~\bibnamefont {Preciado}}, \bibinfo {author} {\bibfnamefont {M.}~\bibnamefont {Isarraraz}}, \bibinfo {author} {\bibfnamefont {I.}~\bibnamefont {Tanabe}}, \bibinfo {author} {\bibfnamefont {T.}~\bibnamefont {Komesu}}, \bibinfo {author} {\bibfnamefont {C.}~\bibnamefont {Troha}}, \bibinfo {author} {\bibfnamefont {L.}~\bibnamefont {Bartels}}, \bibinfo {author} {\bibfnamefont {T.~S.}\ \bibnamefont {Rahman}},\ and\ \bibinfo {author} {\bibfnamefont {P.~A.}\ \bibnamefont {Dowben}},\ }\bibfield  {title} {\bibinfo {title} {Spin–orbit coupling in the band structure of monolayer {W}{$\text{Se}_2$}},\ }\href {https://doi.org/10.1088/0953-8984/27/18/182201} {\bibfield  {journal} {\bibinfo  {journal} {Journal of Physics: Condensed Matter}\ }\textbf {\bibinfo {volume} {27}},\ \bibinfo {pages} {182201} (\bibinfo {year} {2015})}\BibitemShut
  {NoStop}%
\bibitem [{\citenamefont {Alidoust}\ \emph {et~al.}(2014)\citenamefont {Alidoust}, \citenamefont {Bian}, \citenamefont {Xu}, \citenamefont {Sankar}, \citenamefont {Neupane}, \citenamefont {Liu}, \citenamefont {Belopolski}, \citenamefont {Qu}, \citenamefont {Denlinger}, \citenamefont {Chou},\ and\ \citenamefont {Hasan}}]{alidoust2014observation}%
  \BibitemOpen
  \bibfield  {author} {\bibinfo {author} {\bibfnamefont {N.}~\bibnamefont {Alidoust}}, \bibinfo {author} {\bibfnamefont {G.}~\bibnamefont {Bian}}, \bibinfo {author} {\bibfnamefont {S.-Y.}\ \bibnamefont {Xu}}, \bibinfo {author} {\bibfnamefont {R.}~\bibnamefont {Sankar}}, \bibinfo {author} {\bibfnamefont {M.}~\bibnamefont {Neupane}}, \bibinfo {author} {\bibfnamefont {C.}~\bibnamefont {Liu}}, \bibinfo {author} {\bibfnamefont {I.}~\bibnamefont {Belopolski}}, \bibinfo {author} {\bibfnamefont {D.-X.}\ \bibnamefont {Qu}}, \bibinfo {author} {\bibfnamefont {J.~D.}\ \bibnamefont {Denlinger}}, \bibinfo {author} {\bibfnamefont {F.-C.}\ \bibnamefont {Chou}},\ and\ \bibinfo {author} {\bibfnamefont {M.~Z.}\ \bibnamefont {Hasan}},\ }\bibfield  {title} {\bibinfo {title} {Observation of monolayer valence band spin-orbit effect and induced quantum well states in {Mo}{$\text{X}_2$}},\ }\href {https://doi.org/10.1038/ncomms5673} {\bibfield  {journal} {\bibinfo  {journal} {Nature Communications}\ }\textbf {\bibinfo {volume} {5}},\
  \bibinfo {pages} {4673} (\bibinfo {year} {2014})}\BibitemShut {NoStop}%
\bibitem [{\citenamefont {Ciorga}\ \emph {et~al.}(2000)\citenamefont {Ciorga}, \citenamefont {Sachrajda}, \citenamefont {Hawrylak}, \citenamefont {Gould}, \citenamefont {Zawadzki}, \citenamefont {Jullian}, \citenamefont {Feng},\ and\ \citenamefont {Wasilewski}}]{Ciorga2000}%
  \BibitemOpen
  \bibfield  {author} {\bibinfo {author} {\bibfnamefont {M.}~\bibnamefont {Ciorga}}, \bibinfo {author} {\bibfnamefont {A.~S.}\ \bibnamefont {Sachrajda}}, \bibinfo {author} {\bibfnamefont {P.}~\bibnamefont {Hawrylak}}, \bibinfo {author} {\bibfnamefont {C.}~\bibnamefont {Gould}}, \bibinfo {author} {\bibfnamefont {P.}~\bibnamefont {Zawadzki}}, \bibinfo {author} {\bibfnamefont {S.}~\bibnamefont {Jullian}}, \bibinfo {author} {\bibfnamefont {Y.}~\bibnamefont {Feng}},\ and\ \bibinfo {author} {\bibfnamefont {Z.}~\bibnamefont {Wasilewski}},\ }\bibfield  {title} {\bibinfo {title} {Addition spectrum of a lateral dot from coulomb and spin-blockade spectroscopy},\ }\href {https://doi.org/10.1103/PhysRevB.61.R16315} {\bibfield  {journal} {\bibinfo  {journal} {Phys. Rev. B}\ }\textbf {\bibinfo {volume} {61}},\ \bibinfo {pages} {R16315} (\bibinfo {year} {2000})}\BibitemShut {NoStop}%
\bibitem [{\citenamefont {G{\"u}{\c c}l{\"u}}\ \emph {et~al.}(2014)\citenamefont {G{\"u}{\c c}l{\"u}}, \citenamefont {Potasz}, \citenamefont {Korkusinski},\ and\ \citenamefont {Hawrylak}}]{Guclu_Hawrylak_2014}%
  \BibitemOpen
  \bibfield  {author} {\bibinfo {author} {\bibfnamefont {A.~D.}\ \bibnamefont {G{\"u}{\c c}l{\"u}}}, \bibinfo {author} {\bibfnamefont {P.}~\bibnamefont {Potasz}}, \bibinfo {author} {\bibfnamefont {M.}~\bibnamefont {Korkusinski}},\ and\ \bibinfo {author} {\bibfnamefont {P.}~\bibnamefont {Hawrylak}},\ }\href@noop {} {\emph {\bibinfo {title} {{Graphene quantum dots }}}}\ (\bibinfo  {publisher} {Springer},\ \bibinfo {year} {2014})\BibitemShut {NoStop}%
\bibitem [{\citenamefont {G\"uttinger}\ \emph {et~al.}(2010)\citenamefont {G\"uttinger}, \citenamefont {Frey}, \citenamefont {Stampfer}, \citenamefont {Ihn},\ and\ \citenamefont {Ensslin}}]{Guttinger_Ensslin_2010}%
  \BibitemOpen
  \bibfield  {author} {\bibinfo {author} {\bibfnamefont {J.}~\bibnamefont {G\"uttinger}}, \bibinfo {author} {\bibfnamefont {T.}~\bibnamefont {Frey}}, \bibinfo {author} {\bibfnamefont {C.}~\bibnamefont {Stampfer}}, \bibinfo {author} {\bibfnamefont {T.}~\bibnamefont {Ihn}},\ and\ \bibinfo {author} {\bibfnamefont {K.}~\bibnamefont {Ensslin}},\ }\bibfield  {title} {\bibinfo {title} {Spin states in graphene quantum dots},\ }\href {https://doi.org/10.1103/PhysRevLett.105.116801} {\bibfield  {journal} {\bibinfo  {journal} {Phys. Rev. Lett.}\ }\textbf {\bibinfo {volume} {105}},\ \bibinfo {pages} {116801} (\bibinfo {year} {2010})}\BibitemShut {NoStop}%
\bibitem [{\citenamefont {McGuire}(2016)}]{McGuire_2016}%
  \BibitemOpen
  \bibfield  {author} {\bibinfo {author} {\bibfnamefont {J.~A.}\ \bibnamefont {McGuire}},\ }\bibfield  {title} {\bibinfo {title} {Growth and optical properties of colloidal graphene quantum dots},\ }\href {https://doi.org/https://doi.org/10.1002/pssr.201510287} {\bibfield  {journal} {\bibinfo  {journal} {Physica Status Solidi}\ }\textbf {\bibinfo {volume} {10}},\ \bibinfo {pages} {91} (\bibinfo {year} {2016})}\BibitemShut {NoStop}%
\bibitem [{\citenamefont {Wang}\ \emph {et~al.}(2017)\citenamefont {Wang}, \citenamefont {Kharche}, \citenamefont {Costa~Girão}, \citenamefont {Feng}, \citenamefont {M\"{u}llen}, \citenamefont {Meunier}, \citenamefont {Fasel},\ and\ \citenamefont {Ruffieux}}]{Wang_Ruffieux_2017}%
  \BibitemOpen
  \bibfield  {author} {\bibinfo {author} {\bibfnamefont {S.}~\bibnamefont {Wang}}, \bibinfo {author} {\bibfnamefont {N.}~\bibnamefont {Kharche}}, \bibinfo {author} {\bibfnamefont {E.}~\bibnamefont {Costa~Girão}}, \bibinfo {author} {\bibfnamefont {X.}~\bibnamefont {Feng}}, \bibinfo {author} {\bibfnamefont {K.}~\bibnamefont {M\"{u}llen}}, \bibinfo {author} {\bibfnamefont {V.}~\bibnamefont {Meunier}}, \bibinfo {author} {\bibfnamefont {R.}~\bibnamefont {Fasel}},\ and\ \bibinfo {author} {\bibfnamefont {P.}~\bibnamefont {Ruffieux}},\ }\bibfield  {title} {\bibinfo {title} {Quantum dots in graphene nanoribbons},\ }\href {https://doi.org/10.1021/acs.nanolett.7b01244} {\bibfield  {journal} {\bibinfo  {journal} {Nano Letters}\ }\textbf {\bibinfo {volume} {17}},\ \bibinfo {pages} {4277} (\bibinfo {year} {2017})}\BibitemShut {NoStop}%
\bibitem [{\citenamefont {Wang}\ \emph {et~al.}(2018)\citenamefont {Wang}, \citenamefont {De~Greve}, \citenamefont {Jauregui}, \citenamefont {Sushko}, \citenamefont {High}, \citenamefont {Zhou}, \citenamefont {Scuri}, \citenamefont {Taniguchi}, \citenamefont {Watanabe}, \citenamefont {Lukin}, \citenamefont {Park},\ and\ \citenamefont {Kim}}]{Wang_Kim_2018}%
  \BibitemOpen
  \bibfield  {author} {\bibinfo {author} {\bibfnamefont {K.}~\bibnamefont {Wang}}, \bibinfo {author} {\bibfnamefont {K.}~\bibnamefont {De~Greve}}, \bibinfo {author} {\bibfnamefont {L.~A.}\ \bibnamefont {Jauregui}}, \bibinfo {author} {\bibfnamefont {A.}~\bibnamefont {Sushko}}, \bibinfo {author} {\bibfnamefont {A.}~\bibnamefont {High}}, \bibinfo {author} {\bibfnamefont {Y.}~\bibnamefont {Zhou}}, \bibinfo {author} {\bibfnamefont {G.}~\bibnamefont {Scuri}}, \bibinfo {author} {\bibfnamefont {T.}~\bibnamefont {Taniguchi}}, \bibinfo {author} {\bibfnamefont {K.}~\bibnamefont {Watanabe}}, \bibinfo {author} {\bibfnamefont {M.~D.}\ \bibnamefont {Lukin}}, \bibinfo {author} {\bibfnamefont {H.}~\bibnamefont {Park}},\ and\ \bibinfo {author} {\bibfnamefont {P.}~\bibnamefont {Kim}},\ }\bibfield  {title} {\bibinfo {title} {Electrical control of charged carriers and excitons in atomically thin materials},\ }\href {https://doi.org/10.1038/s41565-017-0030-x} {\bibfield  {journal} {\bibinfo  {journal} {Nature Nanotechnology}\
  }\textbf {\bibinfo {volume} {13}},\ \bibinfo {pages} {128} (\bibinfo {year} {2018})}\BibitemShut {NoStop}%
\bibitem [{\citenamefont {Pisoni}\ \emph {et~al.}(2018)\citenamefont {Pisoni}, \citenamefont {Lei}, \citenamefont {Back}, \citenamefont {Eich}, \citenamefont {Overweg}, \citenamefont {Lee}, \citenamefont {Watanabe}, \citenamefont {Taniguchi}, \citenamefont {Ihn},\ and\ \citenamefont {Ensslin}}]{Pisoni_Ensslin_2018}%
  \BibitemOpen
  \bibfield  {author} {\bibinfo {author} {\bibfnamefont {R.}~\bibnamefont {Pisoni}}, \bibinfo {author} {\bibfnamefont {Z.}~\bibnamefont {Lei}}, \bibinfo {author} {\bibfnamefont {P.}~\bibnamefont {Back}}, \bibinfo {author} {\bibfnamefont {M.}~\bibnamefont {Eich}}, \bibinfo {author} {\bibfnamefont {H.}~\bibnamefont {Overweg}}, \bibinfo {author} {\bibfnamefont {Y.}~\bibnamefont {Lee}}, \bibinfo {author} {\bibfnamefont {K.}~\bibnamefont {Watanabe}}, \bibinfo {author} {\bibfnamefont {T.}~\bibnamefont {Taniguchi}}, \bibinfo {author} {\bibfnamefont {T.}~\bibnamefont {Ihn}},\ and\ \bibinfo {author} {\bibfnamefont {K.}~\bibnamefont {Ensslin}},\ }\bibfield  {title} {\bibinfo {title} {{Gate-tunable quantum dot in a high quality single layer {Mo}{$\text{S}_2$} van der Waals heterostructure}},\ }\href {https://doi.org/10.1063/1.5021113} {\bibfield  {journal} {\bibinfo  {journal} {Applied Physics Letters}\ }\textbf {\bibinfo {volume} {112}},\ \bibinfo {pages} {123101} (\bibinfo {year} {2018})}\BibitemShut {NoStop}%
\bibitem [{\citenamefont {Volk}\ \emph {et~al.}(2011)\citenamefont {Volk}, \citenamefont {Fringes}, \citenamefont {Terrés}, \citenamefont {Dauber}, \citenamefont {Engels}, \citenamefont {Trellenkamp},\ and\ \citenamefont {Stampfer}}]{Volk_Stampfer_2011}%
  \BibitemOpen
  \bibfield  {author} {\bibinfo {author} {\bibfnamefont {C.}~\bibnamefont {Volk}}, \bibinfo {author} {\bibfnamefont {S.}~\bibnamefont {Fringes}}, \bibinfo {author} {\bibfnamefont {B.}~\bibnamefont {Terrés}}, \bibinfo {author} {\bibfnamefont {J.}~\bibnamefont {Dauber}}, \bibinfo {author} {\bibfnamefont {S.}~\bibnamefont {Engels}}, \bibinfo {author} {\bibfnamefont {S.}~\bibnamefont {Trellenkamp}},\ and\ \bibinfo {author} {\bibfnamefont {C.}~\bibnamefont {Stampfer}},\ }\bibfield  {title} {\bibinfo {title} {Electronic excited states in bilayer graphene double quantum dots},\ }\href {https://doi.org/10.1021/nl201295s} {\bibfield  {journal} {\bibinfo  {journal} {Nano Letters}\ }\textbf {\bibinfo {volume} {11}},\ \bibinfo {pages} {3581} (\bibinfo {year} {2011})}\BibitemShut {NoStop}%
\bibitem [{\citenamefont {Allen}\ \emph {et~al.}(2012)\citenamefont {Allen}, \citenamefont {Martin},\ and\ \citenamefont {Yacoby}}]{Allen_Yacoby_2012}%
  \BibitemOpen
  \bibfield  {author} {\bibinfo {author} {\bibfnamefont {M.~T.}\ \bibnamefont {Allen}}, \bibinfo {author} {\bibfnamefont {J.}~\bibnamefont {Martin}},\ and\ \bibinfo {author} {\bibfnamefont {A.}~\bibnamefont {Yacoby}},\ }\bibfield  {title} {\bibinfo {title} {Gate-defined quantum confinement in suspended bilayer graphene},\ }\href {https://doi.org/10.1038/ncomms1945} {\bibfield  {journal} {\bibinfo  {journal} {Nature Communications}\ }\textbf {\bibinfo {volume} {3}},\ \bibinfo {pages} {934} (\bibinfo {year} {2012})}\BibitemShut {NoStop}%
\bibitem [{\citenamefont {Eich}\ \emph {et~al.}(2018)\citenamefont {Eich}, \citenamefont {Herman}, \citenamefont {Pisoni}, \citenamefont {Overweg}, \citenamefont {Kurzmann}, \citenamefont {Lee}, \citenamefont {Rickhaus}, \citenamefont {Watanabe}, \citenamefont {Taniguchi}, \citenamefont {Sigrist}, \citenamefont {Ihn},\ and\ \citenamefont {Ensslin}}]{Eich_Ensslin_2018}%
  \BibitemOpen
  \bibfield  {author} {\bibinfo {author} {\bibfnamefont {M.}~\bibnamefont {Eich}}, \bibinfo {author} {\bibfnamefont {F.~c.~v.}\ \bibnamefont {Herman}}, \bibinfo {author} {\bibfnamefont {R.}~\bibnamefont {Pisoni}}, \bibinfo {author} {\bibfnamefont {H.}~\bibnamefont {Overweg}}, \bibinfo {author} {\bibfnamefont {A.}~\bibnamefont {Kurzmann}}, \bibinfo {author} {\bibfnamefont {Y.}~\bibnamefont {Lee}}, \bibinfo {author} {\bibfnamefont {P.}~\bibnamefont {Rickhaus}}, \bibinfo {author} {\bibfnamefont {K.}~\bibnamefont {Watanabe}}, \bibinfo {author} {\bibfnamefont {T.}~\bibnamefont {Taniguchi}}, \bibinfo {author} {\bibfnamefont {M.}~\bibnamefont {Sigrist}}, \bibinfo {author} {\bibfnamefont {T.}~\bibnamefont {Ihn}},\ and\ \bibinfo {author} {\bibfnamefont {K.}~\bibnamefont {Ensslin}},\ }\bibfield  {title} {\bibinfo {title} {Spin and valley states in gate-defined bilayer graphene quantum dots},\ }\href {https://doi.org/10.1103/PhysRevX.8.031023} {\bibfield  {journal} {\bibinfo  {journal} {Phys. Rev. X}\ }\textbf {\bibinfo
  {volume} {8}},\ \bibinfo {pages} {031023} (\bibinfo {year} {2018})}\BibitemShut {NoStop}%
\bibitem [{\citenamefont {Kurzmann}\ \emph {et~al.}(2019)\citenamefont {Kurzmann}, \citenamefont {Eich}, \citenamefont {Overweg}, \citenamefont {Mangold}, \citenamefont {Herman}, \citenamefont {Rickhaus}, \citenamefont {Pisoni}, \citenamefont {Lee}, \citenamefont {Garreis}, \citenamefont {Tong}, \citenamefont {Watanabe}, \citenamefont {Taniguchi}, \citenamefont {Ensslin},\ and\ \citenamefont {Ihn}}]{Kurzmann_Ihn_2019}%
  \BibitemOpen
  \bibfield  {author} {\bibinfo {author} {\bibfnamefont {A.}~\bibnamefont {Kurzmann}}, \bibinfo {author} {\bibfnamefont {M.}~\bibnamefont {Eich}}, \bibinfo {author} {\bibfnamefont {H.}~\bibnamefont {Overweg}}, \bibinfo {author} {\bibfnamefont {M.}~\bibnamefont {Mangold}}, \bibinfo {author} {\bibfnamefont {F.}~\bibnamefont {Herman}}, \bibinfo {author} {\bibfnamefont {P.}~\bibnamefont {Rickhaus}}, \bibinfo {author} {\bibfnamefont {R.}~\bibnamefont {Pisoni}}, \bibinfo {author} {\bibfnamefont {Y.}~\bibnamefont {Lee}}, \bibinfo {author} {\bibfnamefont {R.}~\bibnamefont {Garreis}}, \bibinfo {author} {\bibfnamefont {C.}~\bibnamefont {Tong}}, \bibinfo {author} {\bibfnamefont {K.}~\bibnamefont {Watanabe}}, \bibinfo {author} {\bibfnamefont {T.}~\bibnamefont {Taniguchi}}, \bibinfo {author} {\bibfnamefont {K.}~\bibnamefont {Ensslin}},\ and\ \bibinfo {author} {\bibfnamefont {T.}~\bibnamefont {Ihn}},\ }\bibfield  {title} {\bibinfo {title} {Excited states in bilayer graphene quantum dots},\ }\href
  {https://doi.org/10.1103/PhysRevLett.123.026803} {\bibfield  {journal} {\bibinfo  {journal} {Phys. Rev. Lett.}\ }\textbf {\bibinfo {volume} {123}},\ \bibinfo {pages} {026803} (\bibinfo {year} {2019})}\BibitemShut {NoStop}%
\bibitem [{\citenamefont {Freitag}\ \emph {et~al.}(2016)\citenamefont {Freitag}, \citenamefont {Chizhova}, \citenamefont {Nemes-Incze}, \citenamefont {Woods}, \citenamefont {Gorbachev}, \citenamefont {Cao}, \citenamefont {Geim}, \citenamefont {Novoselov}, \citenamefont {Burgdörfer}, \citenamefont {Libisch},\ and\ \citenamefont {Morgenstern}}]{freitag2016electrostatically}%
  \BibitemOpen
  \bibfield  {author} {\bibinfo {author} {\bibfnamefont {N.~M.}\ \bibnamefont {Freitag}}, \bibinfo {author} {\bibfnamefont {L.~A.}\ \bibnamefont {Chizhova}}, \bibinfo {author} {\bibfnamefont {P.}~\bibnamefont {Nemes-Incze}}, \bibinfo {author} {\bibfnamefont {C.~R.}\ \bibnamefont {Woods}}, \bibinfo {author} {\bibfnamefont {R.~V.}\ \bibnamefont {Gorbachev}}, \bibinfo {author} {\bibfnamefont {Y.}~\bibnamefont {Cao}}, \bibinfo {author} {\bibfnamefont {A.~K.}\ \bibnamefont {Geim}}, \bibinfo {author} {\bibfnamefont {K.~S.}\ \bibnamefont {Novoselov}}, \bibinfo {author} {\bibfnamefont {J.}~\bibnamefont {Burgdörfer}}, \bibinfo {author} {\bibfnamefont {F.}~\bibnamefont {Libisch}},\ and\ \bibinfo {author} {\bibfnamefont {M.}~\bibnamefont {Morgenstern}},\ }\bibfield  {title} {\bibinfo {title} {Electrostatically confined monolayer graphene quantum dots with orbital and valley splittings},\ }\href {https://doi.org/10.1021/acs.nanolett.6b02548} {\bibfield  {journal} {\bibinfo  {journal} {Nano Letters}\ }\textbf {\bibinfo
  {volume} {16}},\ \bibinfo {pages} {5798} (\bibinfo {year} {2016})}\BibitemShut {NoStop}%
\bibitem [{\citenamefont {Saleem}\ \emph {et~al.}(2023)\citenamefont {Saleem}, \citenamefont {Sadecka}, \citenamefont {Korkusinski}, \citenamefont {Miravet}, \citenamefont {Dusko},\ and\ \citenamefont {Hawrylak}}]{YasserGrapheneQD2023}%
  \BibitemOpen
  \bibfield  {author} {\bibinfo {author} {\bibfnamefont {Y.}~\bibnamefont {Saleem}}, \bibinfo {author} {\bibfnamefont {K.}~\bibnamefont {Sadecka}}, \bibinfo {author} {\bibfnamefont {M.}~\bibnamefont {Korkusinski}}, \bibinfo {author} {\bibfnamefont {D.}~\bibnamefont {Miravet}}, \bibinfo {author} {\bibfnamefont {A.}~\bibnamefont {Dusko}},\ and\ \bibinfo {author} {\bibfnamefont {P.}~\bibnamefont {Hawrylak}},\ }\bibfield  {title} {\bibinfo {title} {Theory of excitons in gated bilayer graphene quantum dots},\ }\href {https://doi.org/10.1021/acs.nanolett.3c00406} {\bibfield  {journal} {\bibinfo  {journal} {Nano Letters}\ }\textbf {\bibinfo {volume} {23}},\ \bibinfo {pages} {2998} (\bibinfo {year} {2023})}\BibitemShut {NoStop}%
\bibitem [{\citenamefont {Davari}\ \emph {et~al.}(2020)\citenamefont {Davari}, \citenamefont {Stacy}, \citenamefont {Mercado}, \citenamefont {Tull}, \citenamefont {Basnet}, \citenamefont {Pandey}, \citenamefont {Watanabe}, \citenamefont {Taniguchi}, \citenamefont {Hu},\ and\ \citenamefont {Churchill}}]{Davari2020}%
  \BibitemOpen
  \bibfield  {author} {\bibinfo {author} {\bibfnamefont {S.}~\bibnamefont {Davari}}, \bibinfo {author} {\bibfnamefont {J.}~\bibnamefont {Stacy}}, \bibinfo {author} {\bibfnamefont {A.}~\bibnamefont {Mercado}}, \bibinfo {author} {\bibfnamefont {J.}~\bibnamefont {Tull}}, \bibinfo {author} {\bibfnamefont {R.}~\bibnamefont {Basnet}}, \bibinfo {author} {\bibfnamefont {K.}~\bibnamefont {Pandey}}, \bibinfo {author} {\bibfnamefont {K.}~\bibnamefont {Watanabe}}, \bibinfo {author} {\bibfnamefont {T.}~\bibnamefont {Taniguchi}}, \bibinfo {author} {\bibfnamefont {J.}~\bibnamefont {Hu}},\ and\ \bibinfo {author} {\bibfnamefont {H.}~\bibnamefont {Churchill}},\ }\bibfield  {title} {\bibinfo {title} {Gate-defined accumulation-mode quantum dots in monolayer and bilayer ${\mathrm{w}\mathrm{se}}_{2}$},\ }\href {https://doi.org/10.1103/PhysRevApplied.13.054058} {\bibfield  {journal} {\bibinfo  {journal} {Phys. Rev. Appl.}\ }\textbf {\bibinfo {volume} {13}},\ \bibinfo {pages} {054058} (\bibinfo {year} {2020})}\BibitemShut {NoStop}%
\bibitem [{\citenamefont {Brotons-Gisbert}\ \emph {et~al.}(2019)\citenamefont {Brotons-Gisbert}, \citenamefont {Branny}, \citenamefont {Kumar}, \citenamefont {Picard}, \citenamefont {Proux}, \citenamefont {Gray}, \citenamefont {Burch}, \citenamefont {Watanabe}, \citenamefont {Taniguchi},\ and\ \citenamefont {Gerardot}}]{BrotonsGisbert_Gerardot_2019}%
  \BibitemOpen
  \bibfield  {author} {\bibinfo {author} {\bibfnamefont {M.}~\bibnamefont {Brotons-Gisbert}}, \bibinfo {author} {\bibfnamefont {A.}~\bibnamefont {Branny}}, \bibinfo {author} {\bibfnamefont {S.}~\bibnamefont {Kumar}}, \bibinfo {author} {\bibfnamefont {R.}~\bibnamefont {Picard}}, \bibinfo {author} {\bibfnamefont {R.}~\bibnamefont {Proux}}, \bibinfo {author} {\bibfnamefont {M.}~\bibnamefont {Gray}}, \bibinfo {author} {\bibfnamefont {K.~S.}\ \bibnamefont {Burch}}, \bibinfo {author} {\bibfnamefont {K.}~\bibnamefont {Watanabe}}, \bibinfo {author} {\bibfnamefont {T.}~\bibnamefont {Taniguchi}},\ and\ \bibinfo {author} {\bibfnamefont {B.~D.}\ \bibnamefont {Gerardot}},\ }\bibfield  {title} {\bibinfo {title} {{Coulomb blockade in an atomically thin quantum dot coupled to a tunable Fermi reservoir}},\ }\href {https://doi.org/10.1038/s41565-019-0402-5} {\bibfield  {journal} {\bibinfo  {journal} {Nature Nanotechnology}\ }\textbf {\bibinfo {volume} {14}},\ \bibinfo {pages} {442} (\bibinfo {year} {2019})}\BibitemShut
  {NoStop}%
\bibitem [{\citenamefont {Lu}\ \emph {et~al.}(2019)\citenamefont {Lu}, \citenamefont {Chen}, \citenamefont {Dubey}, \citenamefont {Yao}, \citenamefont {Li}, \citenamefont {Wang}, \citenamefont {Xiong},\ and\ \citenamefont {Srivastava}}]{Lu_Srivastava_2019}%
  \BibitemOpen
  \bibfield  {author} {\bibinfo {author} {\bibfnamefont {X.}~\bibnamefont {Lu}}, \bibinfo {author} {\bibfnamefont {X.}~\bibnamefont {Chen}}, \bibinfo {author} {\bibfnamefont {S.}~\bibnamefont {Dubey}}, \bibinfo {author} {\bibfnamefont {Q.}~\bibnamefont {Yao}}, \bibinfo {author} {\bibfnamefont {W.}~\bibnamefont {Li}}, \bibinfo {author} {\bibfnamefont {X.}~\bibnamefont {Wang}}, \bibinfo {author} {\bibfnamefont {Q.}~\bibnamefont {Xiong}},\ and\ \bibinfo {author} {\bibfnamefont {A.}~\bibnamefont {Srivastava}},\ }\bibfield  {title} {\bibinfo {title} {Optical initialization of a single spin-valley in charged {W}{$\text{Se}_2$} quantum dots},\ }\href {https://doi.org/10.1038/s41565-019-0394-1} {\bibfield  {journal} {\bibinfo  {journal} {Nature Nanotechnology}\ }\textbf {\bibinfo {volume} {14}},\ \bibinfo {pages} {426} (\bibinfo {year} {2019})}\BibitemShut {NoStop}%
\bibitem [{\citenamefont {Chakraborty}\ \emph {et~al.}(2018)\citenamefont {Chakraborty}, \citenamefont {Qiu}, \citenamefont {Konthasinghe}, \citenamefont {Mukherjee}, \citenamefont {Dhara},\ and\ \citenamefont {Vamivakas}}]{Chakraborty_Vamivakas_2018}%
  \BibitemOpen
  \bibfield  {author} {\bibinfo {author} {\bibfnamefont {C.}~\bibnamefont {Chakraborty}}, \bibinfo {author} {\bibfnamefont {L.}~\bibnamefont {Qiu}}, \bibinfo {author} {\bibfnamefont {K.}~\bibnamefont {Konthasinghe}}, \bibinfo {author} {\bibfnamefont {A.}~\bibnamefont {Mukherjee}}, \bibinfo {author} {\bibfnamefont {S.}~\bibnamefont {Dhara}},\ and\ \bibinfo {author} {\bibfnamefont {N.}~\bibnamefont {Vamivakas}},\ }\bibfield  {title} {\bibinfo {title} {{3D Localized Trions in Monolayer {W}{$\text{Se}_2$} in a Charge Tunable van der Waals Heterostructure}},\ }\href {https://doi.org/10.1021/acs.nanolett.7b05409} {\bibfield  {journal} {\bibinfo  {journal} {Nano Letters}\ }\textbf {\bibinfo {volume} {18}},\ \bibinfo {pages} {2859} (\bibinfo {year} {2018})}\BibitemShut {NoStop}%
\bibitem [{\citenamefont {Zhang}\ \emph {et~al.}(2017)\citenamefont {Zhang}, \citenamefont {Song}, \citenamefont {Luo}, \citenamefont {Deng}, \citenamefont {Mosallanejad}, \citenamefont {Taniguchi}, \citenamefont {Watanabe}, \citenamefont {Li}, \citenamefont {Cao}, \citenamefont {Guo}, \citenamefont {Nori},\ and\ \citenamefont {Guo}}]{Zhang_Guo_2017}%
  \BibitemOpen
  \bibfield  {author} {\bibinfo {author} {\bibfnamefont {Z.-Z.}\ \bibnamefont {Zhang}}, \bibinfo {author} {\bibfnamefont {X.-X.}\ \bibnamefont {Song}}, \bibinfo {author} {\bibfnamefont {G.}~\bibnamefont {Luo}}, \bibinfo {author} {\bibfnamefont {G.-W.}\ \bibnamefont {Deng}}, \bibinfo {author} {\bibfnamefont {V.}~\bibnamefont {Mosallanejad}}, \bibinfo {author} {\bibfnamefont {T.}~\bibnamefont {Taniguchi}}, \bibinfo {author} {\bibfnamefont {K.}~\bibnamefont {Watanabe}}, \bibinfo {author} {\bibfnamefont {H.-O.}\ \bibnamefont {Li}}, \bibinfo {author} {\bibfnamefont {G.}~\bibnamefont {Cao}}, \bibinfo {author} {\bibfnamefont {G.-C.}\ \bibnamefont {Guo}}, \bibinfo {author} {\bibfnamefont {F.}~\bibnamefont {Nori}},\ and\ \bibinfo {author} {\bibfnamefont {G.-P.}\ \bibnamefont {Guo}},\ }\bibfield  {title} {\bibinfo {title} {{Electrotunable artificial molecules based on van der Waals heterostructures}},\ }\href {https://doi.org/10.1126/sciadv.1701699} {\bibfield  {journal} {\bibinfo  {journal} {Science Advances}\
  }\textbf {\bibinfo {volume} {3}},\ \bibinfo {pages} {e1701699} (\bibinfo {year} {2017})}\BibitemShut {NoStop}%
\bibitem [{\citenamefont {Bhandari}\ \emph {et~al.}(2018)\citenamefont {Bhandari}, \citenamefont {Wang}, \citenamefont {Watanabe}, \citenamefont {Taniguchi}, \citenamefont {Kim},\ and\ \citenamefont {Westervelt}}]{bhandari2018imaging}%
  \BibitemOpen
  \bibfield  {author} {\bibinfo {author} {\bibfnamefont {S.}~\bibnamefont {Bhandari}}, \bibinfo {author} {\bibfnamefont {K.}~\bibnamefont {Wang}}, \bibinfo {author} {\bibfnamefont {K.}~\bibnamefont {Watanabe}}, \bibinfo {author} {\bibfnamefont {T.}~\bibnamefont {Taniguchi}}, \bibinfo {author} {\bibfnamefont {P.}~\bibnamefont {Kim}},\ and\ \bibinfo {author} {\bibfnamefont {R.}~\bibnamefont {Westervelt}},\ }\bibfield  {title} {\bibinfo {title} {Imaging quantum dot formation in {Mo}{$\text{S}_2$} nanostructures},\ }\href {https://doi.org/10.1088/1361-6528/aad79f} {\bibfield  {journal} {\bibinfo  {journal} {Nanotechnology}\ }\textbf {\bibinfo {volume} {29}},\ \bibinfo {pages} {42LT03} (\bibinfo {year} {2018})}\BibitemShut {NoStop}%
\bibitem [{\citenamefont {Chen}\ \emph {et~al.}(2018)\citenamefont {Chen}, \citenamefont {Li},\ and\ \citenamefont {Peeters}}]{chen2018magnetic}%
  \BibitemOpen
  \bibfield  {author} {\bibinfo {author} {\bibfnamefont {Q.}~\bibnamefont {Chen}}, \bibinfo {author} {\bibfnamefont {L.}~\bibnamefont {Li}},\ and\ \bibinfo {author} {\bibfnamefont {F.}~\bibnamefont {Peeters}},\ }\bibfield  {title} {\bibinfo {title} {Magnetic field dependence of electronic properties of {Mo}{$\text{S}_2$} quantum dots with different edges},\ }\href {https://doi.org/10.1103/PhysRevB.97.085437} {\bibfield  {journal} {\bibinfo  {journal} {Physical Review B}\ }\textbf {\bibinfo {volume} {97}},\ \bibinfo {pages} {085437} (\bibinfo {year} {2018})}\BibitemShut {NoStop}%
\bibitem [{\citenamefont {Szulakowska}\ \emph {et~al.}(2020)\citenamefont {Szulakowska}, \citenamefont {Cygorek}, \citenamefont {Bieniek},\ and\ \citenamefont {Hawrylak}}]{LudkaQD2020}%
  \BibitemOpen
  \bibfield  {author} {\bibinfo {author} {\bibfnamefont {L.}~\bibnamefont {Szulakowska}}, \bibinfo {author} {\bibfnamefont {M.}~\bibnamefont {Cygorek}}, \bibinfo {author} {\bibfnamefont {M.}~\bibnamefont {Bieniek}},\ and\ \bibinfo {author} {\bibfnamefont {P.}~\bibnamefont {Hawrylak}},\ }\bibfield  {title} {\bibinfo {title} {Valley- and spin-polarized broken-symmetry states of interacting electrons in gated {Mo}{$\text{S}_2$} quantum dots},\ }\href {https://doi.org/10.1103/PhysRevB.102.245410} {\bibfield  {journal} {\bibinfo  {journal} {Phys. Rev. B}\ }\textbf {\bibinfo {volume} {102}},\ \bibinfo {pages} {245410} (\bibinfo {year} {2020})}\BibitemShut {NoStop}%
\bibitem [{\citenamefont {Paw\l{}owski}\ \emph {et~al.}(2021)\citenamefont {Paw\l{}owski}, \citenamefont {Bieniek},\ and\ \citenamefont {Wo\ifmmode~\acute{z}\else \'{z}\fi{}niak}}]{JarekPRA2021}%
  \BibitemOpen
  \bibfield  {author} {\bibinfo {author} {\bibfnamefont {J.}~\bibnamefont {Paw\l{}owski}}, \bibinfo {author} {\bibfnamefont {M.}~\bibnamefont {Bieniek}},\ and\ \bibinfo {author} {\bibfnamefont {T.}~\bibnamefont {Wo\ifmmode~\acute{z}\else \'{z}\fi{}niak}},\ }\bibfield  {title} {\bibinfo {title} {Valley two-qubit system in a {Mo}{$\text{S}_2$}-monolayer gated double quantum dot},\ }\href {https://doi.org/10.1103/PhysRevApplied.15.054025} {\bibfield  {journal} {\bibinfo  {journal} {Phys. Rev. Appl.}\ }\textbf {\bibinfo {volume} {15}},\ \bibinfo {pages} {054025} (\bibinfo {year} {2021})}\BibitemShut {NoStop}%
\bibitem [{\citenamefont {Korkusinski}\ \emph {et~al.}(2023)\citenamefont {Korkusinski}, \citenamefont {Saleem}, \citenamefont {Dusko}, \citenamefont {Miravet},\ and\ \citenamefont {Hawrylak}}]{Marek_nanoletters_2023}%
  \BibitemOpen
  \bibfield  {author} {\bibinfo {author} {\bibfnamefont {M.}~\bibnamefont {Korkusinski}}, \bibinfo {author} {\bibfnamefont {Y.}~\bibnamefont {Saleem}}, \bibinfo {author} {\bibfnamefont {A.}~\bibnamefont {Dusko}}, \bibinfo {author} {\bibfnamefont {D.}~\bibnamefont {Miravet}},\ and\ \bibinfo {author} {\bibfnamefont {P.}~\bibnamefont {Hawrylak}},\ }\bibfield  {title} {\bibinfo {title} {Spontaneous spin and valley symmetry-broken states of interacting massive dirac fermions in a bilayer graphene quantum dot},\ }\href {https://doi.org/10.1021/acs.nanolett.3c02073} {\bibfield  {journal} {\bibinfo  {journal} {Nano Letters}\ }\textbf {\bibinfo {volume} {23}},\ \bibinfo {pages} {7546} (\bibinfo {year} {2023})},\ \bibinfo {note} {pMID: 37561956},\ \Eprint {https://arxiv.org/abs/https://doi.org/10.1021/acs.nanolett.3c02073} {https://doi.org/10.1021/acs.nanolett.3c02073} \BibitemShut {NoStop}%
\bibitem [{\citenamefont {Pawłowski}\ \emph {et~al.}(2024)\citenamefont {Pawłowski}, \citenamefont {Miravet}, \citenamefont {Bieniek}, \citenamefont {Korkusinski}, \citenamefont {Boddison-Chouinard}, \citenamefont {Gaudreau}, \citenamefont {Luican-Mayer},\ and\ \citenamefont {Hawrylak}}]{pawłowski2024interacting}%
  \BibitemOpen
  \bibfield  {author} {\bibinfo {author} {\bibfnamefont {J.}~\bibnamefont {Pawłowski}}, \bibinfo {author} {\bibfnamefont {D.}~\bibnamefont {Miravet}}, \bibinfo {author} {\bibfnamefont {M.}~\bibnamefont {Bieniek}}, \bibinfo {author} {\bibfnamefont {M.}~\bibnamefont {Korkusinski}}, \bibinfo {author} {\bibfnamefont {J.}~\bibnamefont {Boddison-Chouinard}}, \bibinfo {author} {\bibfnamefont {L.}~\bibnamefont {Gaudreau}}, \bibinfo {author} {\bibfnamefont {A.}~\bibnamefont {Luican-Mayer}},\ and\ \bibinfo {author} {\bibfnamefont {P.}~\bibnamefont {Hawrylak}},\ }\href@noop {} {\bibinfo {title} {Interacting holes in a gated {W}{$\text{Se}_2$} quantum channel: valley correlations and zigzag wigner crystal}} (\bibinfo {year} {2024}),\ \Eprint {https://arxiv.org/abs/2406.08655} {arXiv:2406.08655} \BibitemShut {NoStop}%
\bibitem [{\citenamefont {He}\ \emph {et~al.}(2014)\citenamefont {He}, \citenamefont {Kumar}, \citenamefont {Zhao}, \citenamefont {Wang}, \citenamefont {Mak}, \citenamefont {Zhao},\ and\ \citenamefont {Shan}}]{HePRL2014}%
  \BibitemOpen
  \bibfield  {author} {\bibinfo {author} {\bibfnamefont {K.}~\bibnamefont {He}}, \bibinfo {author} {\bibfnamefont {N.}~\bibnamefont {Kumar}}, \bibinfo {author} {\bibfnamefont {L.}~\bibnamefont {Zhao}}, \bibinfo {author} {\bibfnamefont {Z.}~\bibnamefont {Wang}}, \bibinfo {author} {\bibfnamefont {K.~F.}\ \bibnamefont {Mak}}, \bibinfo {author} {\bibfnamefont {H.}~\bibnamefont {Zhao}},\ and\ \bibinfo {author} {\bibfnamefont {J.}~\bibnamefont {Shan}},\ }\bibfield  {title} {\bibinfo {title} {Tightly bound excitons in monolayer ${\mathrm{wse}}_{2}$},\ }\href {https://doi.org/10.1103/PhysRevLett.113.026803} {\bibfield  {journal} {\bibinfo  {journal} {Phys. Rev. Lett.}\ }\textbf {\bibinfo {volume} {113}},\ \bibinfo {pages} {026803} (\bibinfo {year} {2014})}\BibitemShut {NoStop}%
\bibitem [{\citenamefont {Wang}\ \emph {et~al.}(2015)\citenamefont {Wang}, \citenamefont {Marie}, \citenamefont {Gerber}, \citenamefont {Amand}, \citenamefont {Lagarde}, \citenamefont {Bouet}, \citenamefont {Vidal}, \citenamefont {Balocchi},\ and\ \citenamefont {Urbaszek}}]{WangPRL2015}%
  \BibitemOpen
  \bibfield  {author} {\bibinfo {author} {\bibfnamefont {G.}~\bibnamefont {Wang}}, \bibinfo {author} {\bibfnamefont {X.}~\bibnamefont {Marie}}, \bibinfo {author} {\bibfnamefont {I.}~\bibnamefont {Gerber}}, \bibinfo {author} {\bibfnamefont {T.}~\bibnamefont {Amand}}, \bibinfo {author} {\bibfnamefont {D.}~\bibnamefont {Lagarde}}, \bibinfo {author} {\bibfnamefont {L.}~\bibnamefont {Bouet}}, \bibinfo {author} {\bibfnamefont {M.}~\bibnamefont {Vidal}}, \bibinfo {author} {\bibfnamefont {A.}~\bibnamefont {Balocchi}},\ and\ \bibinfo {author} {\bibfnamefont {B.}~\bibnamefont {Urbaszek}},\ }\bibfield  {title} {\bibinfo {title} {Giant enhancement of the optical second-harmonic emission of ${\mathrm{wse}}_{2}$ monolayers by laser excitation at exciton resonances},\ }\href {https://doi.org/10.1103/PhysRevLett.114.097403} {\bibfield  {journal} {\bibinfo  {journal} {Phys. Rev. Lett.}\ }\textbf {\bibinfo {volume} {114}},\ \bibinfo {pages} {097403} (\bibinfo {year} {2015})}\BibitemShut {NoStop}%
\bibitem [{\citenamefont {Song}\ \emph {et~al.}(2015)\citenamefont {Song}, \citenamefont {Liu}, \citenamefont {Mosallanejad}, \citenamefont {You}, \citenamefont {Han}, \citenamefont {Chen}, \citenamefont {Li}, \citenamefont {Cao}, \citenamefont {Xiao}, \citenamefont {Guo},\ and\ \citenamefont {Guo}}]{song2015gate}%
  \BibitemOpen
  \bibfield  {author} {\bibinfo {author} {\bibfnamefont {X.-X.}\ \bibnamefont {Song}}, \bibinfo {author} {\bibfnamefont {D.}~\bibnamefont {Liu}}, \bibinfo {author} {\bibfnamefont {V.}~\bibnamefont {Mosallanejad}}, \bibinfo {author} {\bibfnamefont {J.}~\bibnamefont {You}}, \bibinfo {author} {\bibfnamefont {T.-Y.}\ \bibnamefont {Han}}, \bibinfo {author} {\bibfnamefont {D.-T.}\ \bibnamefont {Chen}}, \bibinfo {author} {\bibfnamefont {H.-O.}\ \bibnamefont {Li}}, \bibinfo {author} {\bibfnamefont {G.}~\bibnamefont {Cao}}, \bibinfo {author} {\bibfnamefont {M.}~\bibnamefont {Xiao}}, \bibinfo {author} {\bibfnamefont {G.-C.}\ \bibnamefont {Guo}},\ and\ \bibinfo {author} {\bibfnamefont {G.-P.}\ \bibnamefont {Guo}},\ }\bibfield  {title} {\bibinfo {title} {A gate defined quantum dot on the two-dimensional transition metal dichalcogenide semiconductor {W}{$\text{Se}_2$}},\ }\href {https://doi.org/10.1039/C5NR04961J} {\bibfield  {journal} {\bibinfo  {journal} {Nanoscale}\ }\textbf {\bibinfo {volume} {7}},\ \bibinfo {pages}
  {16867} (\bibinfo {year} {2015})}\BibitemShut {NoStop}%
\bibitem [{\citenamefont {Janssens}\ \emph {et~al.}(2001)\citenamefont {Janssens}, \citenamefont {Partoens},\ and\ \citenamefont {Peeters}}]{IIexcitonsPeeters2001}%
  \BibitemOpen
  \bibfield  {author} {\bibinfo {author} {\bibfnamefont {K.~L.}\ \bibnamefont {Janssens}}, \bibinfo {author} {\bibfnamefont {B.}~\bibnamefont {Partoens}},\ and\ \bibinfo {author} {\bibfnamefont {F.~M.}\ \bibnamefont {Peeters}},\ }\bibfield  {title} {\bibinfo {title} {Magnetoexcitons in planar type-ii quantum dots in a perpendicular magnetic field},\ }\href {https://doi.org/10.1103/PhysRevB.64.155324} {\bibfield  {journal} {\bibinfo  {journal} {Phys. Rev. B}\ }\textbf {\bibinfo {volume} {64}},\ \bibinfo {pages} {155324} (\bibinfo {year} {2001})}\BibitemShut {NoStop}%
\end{thebibliography}%
%%%%%%%%%%%%%%%%%%%%%%%%%%%%%%%%%%%%%%%%%%%%%%%%%%%%%%%%%%%%%%%%%%%%%%%%%%%%%%%%%%%%%%%%%%%%%%%%

\end{document}